# The Topology of Canonical Flux Tubes in Flared Jet Geometry

Eric Sander Lavine    Setthivoine You

*University of Washington 4000 15th Street NE Aeronautics and Astronautics 211 Guggenheim Hall, Box 352400 Seattle, WA 98195 USA; slavine2@uw.edu, syou@aa.washington.edu*

**Abstract** Magnetized plasma jets are generally modeled as magnetic flux tubes filled with flowing plasma governed by magnetohydrodynamics (MHD). We outline here a more fundamental approach based on flux tubes of canonical vorticity, where canonical vorticity is defined as the circulation of the species' canonical momentum. This approach extends the concept of magnetic flux tube evolution to include the effects of finite particle momentum and enables visualization of the topology of plasma jets in regimes beyond MHD. A flared, current-carrying magnetic flux tube in an ion-electron plasma with finite ion momentum is thus equivalent to either a pair of electron and ion flow flux tubes, a pair of electron and ion canonical momentum flux tubes, or a pair of electron and ion canonical vorticity flux tubes. We examine the morphology of all these flux tubes for increasing electrical currents, different radial current profiles, different electron Mach numbers, and a fixed, flared, axisymmetric magnetic geometry. Calculations of gauge-invariant relative canonical helicities track the evolution of magnetic, cross, and kinetic helicities in the system, and show that ion flow fields can unwind to compensate for an increasing magnetic twist. The results demonstrate that including a species' finite momentum can result in a very long collimated canonical vorticity flux tube even if the magnetic flux tube is flared. With finite momentum, particle density gradients must be normal to canonical vorticities, not to magnetic fields, so observations of collimated astrophysical jets could be images of canonical vorticity flux tubes instead of magnetic flux tubes.

1. **INTRODUCTION**

Astrophysical jets (De Young 1991) are collimated bipolar high-speed outflows observed to be natural features of objects that spin and accrete matter ranging in scales from protoplanetary nebula (Sahai et al. 2003), young stellar objects (Burrows et al. 1996) to active galactic nuclei (Nakamura et al. 2001). Although source energies and scale lengths vary over many orders of magnitude, common features suggest that universal mechanisms may be responsible for jet launching, collimation, and stability. Hydrodynamic models suppose that polar funnels in a thick accretion disk behave as nozzles for launching a jet outflow (Icke et al. 1992) and collimation beyond the formation region may be preserved for high Mach number flows ($M_{jet} > 10$) by reducing hydrodynamic Kelvin-Helmholtz instabilities, sharpening the end tips, and minimizing mixing with ambient material along the length of the jet (Norman et al. 1982). These hydrodynamic models however, have difficulty explaining the collimation of very long jets observed from low aspect-ratio funnels (De Young 1991), over-dense jets surrounded by vacuum or low-density medium (A. Frank et al. 1998), signatures of helical motion (Broderick & Loeb 2009) or helical magnetic fields (Donati et al. 2006). Magnetohydrodynamic (MHD) models thus include the presence and effect of magnetic fields for launching jets from ionized accretion disks spinning around a central object with a background poloidal magnetic field (Lynden-Bell, 2003), and an MHD "gobble" mechanism in a flared screw pinch configuration that convects magnetic flux from the central engine to pile up at the end tips of the jet and increase the collimation (Bellan 2003, Bellan et al. 2005). Experimental evidence for the "gobble" collimation mechanism has been observed in magnetized laboratory plasma jets produced by planar plasma guns designed to mimic the accretion disk linked to a central object, attaining collimated aspect-ratios of ~10: 1 before going kink unstable (You et al. 2005, Yun et al 2007). Axial shear flows were predicted to stabilize current and pressure-driven instabilities in a Z-pinch configuration (Shumlak & Hartman 1995), and laboratory observations confirmed the theory by producing stable collimated jets with aspect ratios >30:1 with fixed length, fixed end-points, and no poloidal magnetic field (Shumlak et al. 2003). Both of these models are based on large electrical currents flowing in the jet, raising the question as to whether both magnetic collimation and flow stabilization mechanisms could work together to produce high aspect-ratio jets with only one end tied to the central object. Our experiment (Mochi LabJet), currently in the commissioning phase, is designed to explore this possibility by producing sheared helical flows inside a driven, magnetized plasma jet launched from planar, concentric, azimuthally symmetric electrodes.

To account for finite momentum and effects beyond MHD phenomena, we describe the system with the canonical momentum of each species. In ideal MHD, particle excursions off magnetic flux surfaces are neglected, so there is a one-to-one relationship between magnetic flux tubes and the plasma. It is therefore sufficient to track the shape and evolution of magnetic flux tubes. Non-ideal effects like Hall terms, pressure tensors, finite Larmor radius effects, kinetics, and inertia are generally assumed to be localized to distinct control volumes and treated separately as the need arises. These more realistic effects are infrequently treated ab initio in numerical simulations on a global scale because of the computational cost and complexity. Considering a more fundamental approach with a species' canonical momentum preserves such non-ideal effects while allowing us to track the evolution and shape of the plasma with concepts of generalized flux tubes that can change shape. These more realistic effects can then be interpreted as generalized (i.e. canonical) flux tubes reconnecting, twisting, linking, unwinding, etc. under some topological constraint, such as generalized (i.e. canonical) helicity continuity (You 2012, 2014). For example, a cylindrical magnetic flux tube could untwist to twist up an ion kinetic flux tube while preserving the total sum of helicities, or vice versa. The winding of an ion kinetic flux tube is manifested as increasing helical flows in a jet. Conversely, decreasing helical flows (i.e. more axial flows) would accompany increasing magnetic twist in a jet. This approach also unifies multi-species neutral hydrodynamics and plasma dynamics as two opposite limits of the dynamics of canonical momenta. Time- and spatially varying charge-to-mass ratios during the evolution of the canonical flux tubes provides a continuous transition between the two limits. This approach also permits further generalizations to relativistic and gravitational regimes (You 2016) without having to drop our intuition of the geometry of flux tubes.

This paper examines the fields and topology of canonical flux tubes with a fixed, flared, axisymmetric magnetic field for varying electrical current strengths and radial profiles. These geometric and boundary conditions were chosen to be relevant to planar plasma gun laboratory setups designed to mimic an accretion disk-astrophysical jet system. (Bellan et al. 2005; You et al. 2005). These calculations intend to provide an intuitive, kinematic view of the evolution of jets without resorting to costly fully self-consistent dynamical calculations. The electrons are here considered to be massless only to simplify the picture to just two flux tubes (one ion and one electron, equivalent to a magnetic flux tube) but can readily be extended for the more general case of three or more flux tubes (ions, electrons, and magnetic tubes). Section 2 reviews the concepts of canonical momentum, canonical vorticity, and canonical helicity; defines canonical flux tubes; and discusses the governing equations of motion. Section 3 applies these concepts to a fixed, flared, axisymmetric jet in a numerical visualization tool. Section 4 discusses the resulting morphology of the canonical flux tubes in relation to the more familiar current-carrying magnetic flux tube. Section 5 presents the evolution of all the forms of helicities for the cases of Section 4, before concluding in Section 6.

## 2. CANONICAL FLUX TUBES, HELICITIES, AND GOVERNING EQUATIONS

Each species $\sigma$ in a plasma has a canonical momentum $\boldsymbol{P}_\sigma = \rho_\sigma \boldsymbol{u}_\sigma + \rho_{c\sigma} \boldsymbol{A}$, expressed as the weighted sum of the species velocity $\boldsymbol{u}_\sigma$ and the magnetic vector potential $\boldsymbol{A}$. The weights are the mass functions $\rho_\sigma$ and the charge functions $\rho_{c\sigma}$ of the species. For a relativistic single particle of mass $m_\sigma$, charge $q_\sigma$ and velocity $\boldsymbol{v}_\sigma$, these functions reduce to $\rho_\sigma \to \gamma(\boldsymbol{v}_\sigma) m_\sigma$, $\rho_{c\sigma} \to q_\sigma$, and $\boldsymbol{u}_\sigma \to \boldsymbol{v}_\sigma$, where $\gamma(\boldsymbol{v}_\sigma)$ is the Lorentz factor. In kinetic regimes described by the velocity distribution function $f_\sigma$, the functions can be written as $\rho_\sigma \to f_\sigma \gamma m_\sigma$, $\rho_{c\sigma} \to f_\sigma q_\sigma$, and $\boldsymbol{u}_\sigma \to \boldsymbol{v}_\sigma$, where the velocities are system parameters. In fluid regimes described by a number density $n_\sigma$ flowing at the bulk velocity $\boldsymbol{u}_\sigma$, the functions can be written as $\rho_\sigma \to n_\sigma \gamma(\boldsymbol{u}_\sigma) m_\sigma$ and $\rho_{c\sigma} \to n_\sigma q_\sigma$ where the fluid velocity is permitted to be relativistic. Here, we consider a barotropic, non-relativistic, reduced two-fluid system of ions ($\sigma \to i$) and electrons ($\sigma \to e$) so the functions reduce to constants, $\rho_\sigma \to m_\sigma$ and $\rho_{c\sigma} \to q_\sigma$. You (2016) derives and describes in detail the more general cases.

The species' canonical vorticity is defined as the circulation of the canonical momentum $\boldsymbol{\Omega}_\sigma = \boldsymbol{\nabla} \times \boldsymbol{P}_\sigma$. Since here we are restricting the analysis to a reduced two-fluid, $\boldsymbol{\Omega}_\sigma = m_\sigma \boldsymbol{\omega}_\sigma + q_\sigma \boldsymbol{B}$, where the fluid vorticity is $\boldsymbol{\omega}_\sigma = \boldsymbol{\nabla} \times \boldsymbol{u}_\sigma$ and the magnetic field is $\boldsymbol{B} = \boldsymbol{\nabla} \times \boldsymbol{A}$. You (2016) shows the forms of canonical vorticity in the more general cases where gradients of density or distribution function play a significant role. The canonical vorticity flux is defined as the integral of the canonical vorticity over the area enclosed by an arbitrary loop,

$$\Psi_\sigma \equiv \oiint_S \boldsymbol{\Omega}_\sigma \cdot d\boldsymbol{S} = m_\sigma \mathcal{F}_\sigma + q_\sigma \psi \tag{1}$$

where $\mathcal{F}_\sigma \equiv \oiint_S \boldsymbol{\omega}_\sigma \cdot d\boldsymbol{S}$ is defined as the flow vorticity flux and $\psi \equiv \oiint_S \boldsymbol{B} \cdot d\boldsymbol{S}$ is defined as the magnetic flux, both enclosed by the same loop contouring the surface $S$. In the case of an axisymmetric system, the canonical vorticity flux can be directly related to the particle canonical angular momentum in the azimuthal direction $\hat{\theta}$ at a given radial position $r$

$$\Psi_\sigma = 2\pi r P_{\sigma\theta}. \tag{2}$$

Since the particle motion preserves angular momentum, surfaces of constant canonical vorticity flux represent particle drift surfaces. Therefore, tubes of constant canonical vorticity flux (canonical flux tubes for short) replace magnetic flux tubes as the primary topological building block of multi-fluid or kinetic plasmas, extending flux tube physics to include the effects of finite species momentum. A canonical flux tube represents the linear combination of a magnetic flux tube and a flow vorticity flux tube. Section 3 will show graphical representations of these canonical flux tubes for several scenarios. Similar to magnetic flux tubes, there are two main reasons why canonical flux tubes are useful concepts: (1) the evolution of the plasma can be tracked by following the evolution of canonical flux tubes, with the added advantage of being more general than magnetic flux tubes; and (2) concepts of self-organization in magnetostatic plasmas (Taylor 1986) can be extended to the constrained relaxation of flowing magnetized plasmas (Steinhauer & Ishida 1998). We consider these two reasons in turn.

First, the evolution of the plasma is described by the canonical equation of motion written in the form of an Ohm's law (You 2012, 2016)

$$\boldsymbol{\Sigma}_\sigma + \boldsymbol{u}_\sigma \times \boldsymbol{\Omega}_\sigma = -\boldsymbol{R}_\sigma \tag{3}$$

where the canonical force-field is defined as $\boldsymbol{\Sigma}_\sigma \equiv -\nabla h_\sigma - \partial \boldsymbol{P}_\sigma/\partial t$ with an enthalpy $h_\sigma$ which represents conservative and inductive forces; the $\boldsymbol{u}_\sigma \times \boldsymbol{\Omega}_\sigma$ term represents forces that do no work (Coriolis and Lorentz forces); and $\boldsymbol{R}_\sigma$ amalgamates all the dissipative forces. You (2016) demonstrates from a Lagrangian-Hamiltonian formalism that equation (3) is valid in single particle regimes, kinetic regimes, and fluid regimes, at classical and relativistic scales, provided the appropriate definitions of canonical momentum and enthalpy are chosen (You 2016 – Table 1). For example, the Vlasov-Boltzmann equation can be derived from equation (3) if the enthalpy of a kinetic distribution $f_\sigma$ is $h_\sigma = f_\sigma(\gamma m_\sigma c^2 + q_\sigma \phi)$. In our two-fluid classical case, the fluid equations can be derived if the enthalpy is $h_\sigma = \rho_\sigma u_\sigma^2/2 + \rho_{c\sigma}\phi + \int dp_\sigma$, combining the conservative potentials (electrostatic $\phi$ and scalar pressure $dp_\sigma$) with the kinetic energy. You (2016) also demonstrates an isomorphism between equation (3) and a general form of Maxwell's equations for the canonical fields $\boldsymbol{\Sigma}_\sigma$ and $\boldsymbol{\Omega}_\sigma$, so for any regime, the canonical vorticity and the force-field evolve according to Maxwell's equations, and the enthalpy and canonical momentum act as scalar and vector potentials for these fields. Taking the circulation of equation (3) in the absence of friction gives the induction equation

$$\frac{\partial \boldsymbol{\Omega}_\sigma}{\partial t} - \nabla \times (\boldsymbol{u}_\sigma \times \boldsymbol{\Omega}_\sigma) = 0 \qquad (4)$$

and shows that the canonical vorticity $\boldsymbol{\Omega}_\sigma$ is frozen to the species motion in any regime, not only the fluid regime. There is therefore a one-to-one relationship between the evolution of the canonical flux tubes and the evolution of the species. If we ignore the species' inertia ($m_\sigma \to 0$), then the induction equation reduces to the usual frozen-in magnetic flux condition, where there is a one-to-one relationship between the evolution of the magnetic flux tube and the evolution of the plasma. Dotting equation (3) with $\boldsymbol{\Omega}_\sigma$ gives

$$\nabla h_\sigma \cdot \boldsymbol{\Omega}_\sigma = 0 \qquad (5)$$

in steady-state and the absence of friction. This condition is the generalization of the magnetostatic $\nabla p_\sigma \cdot \boldsymbol{B} = 0$ where the pressure gradient is always perpendicular to the magnetic flux surfaces. Supposing that light emission is proportional to (the square of) particle density in a uniform temperature ideal magnetostatic plasma, photographs would then be signatures of magnetic flux surfaces. In the more general case of a flowing plasma where enthalpy is purely density driven (so equation (5) becomes $\nabla n \cdot \boldsymbol{\Omega}_\sigma = 0$), then photographs would be signatures of canonical vorticity flux surfaces not magnetic flux surfaces.

Second, plasmas often appear to self-organize into certain preferred states, independent of the detailed dynamics but subject to constrained relaxation (Brown 1997). For example, global plasma behavior is thought to be constrained by the conservation of magnetic helicity $\mathcal{K} \equiv \int \boldsymbol{A} \cdot \boldsymbol{B} \, dV$ (Taylor 1974, Taylor 1986), a measurement of the twist, writhes, and links of magnetic flux tubes, provided the assumptions behind ideal MHD are satisfied everywhere except within small reconnection regions. Because these assumptions freeze plasma to the magnetic flux, from a topological point of view there is no need to distinguish between the behavior of the plasma fluid and the shape of the magnetic fields. Relaxation theories based on these concepts have been successful at predicting the self-organization of certain classes of magnetofluids into force-free states and visualizing the kinematic evolution of magnetic flux tubes (Bellan 2000). However, plasmas are not generally force free, and may exhibit significant fluid pressures and flows, so a model based on a more general helicity such as the species' canonical helicity $K_\sigma \equiv \int \boldsymbol{P}_\sigma \cdot \boldsymbol{\Omega}_\sigma \, dV$ provides a more fundamental approach to helicity-constrained relaxation. Such a model (Steinhauer & Ishida 1997, You 2012) retrieves all the previous results for static ideal MHD plasmas (Taylor 1974, Finn & Antonsen 1985, Woltjer 1986, Ji 1999) and ordinary neutral fluids (Moffat 1969) in the appropriate limits, while extending the same topological concepts to regimes where flowing multi-fluid plasmas are applicable (Steinhauer & Ishida 1997, Steinhauer & Ishida 1998, Steinhauer et al. 2001).

In a closed volume, canonical helicity is gauge independent and has a well-defined value only when the canonical vorticity at the boundary is purely tangential, i.e., if $\mathbf{\Omega}_\sigma \cdot \hat{\mathbf{n}}|_s = 0$ where $\hat{\mathbf{n}}$ here represents the normal unit vector to a surface $S$ bounding the closed volume. Early generalizations of helicity (Turner 1986, Avinash 1992, Oliveira & Tajima 1995, Steinhauer & Ishida 1997) therefore, ignored situations where canonical flux tubes intercept the boundaries of a system and, even within an isolated system, ignored that each species' canonical flux tubes can overlap and intercept one another. These treatments thus concluded that each species' canonical helicity was itself invariant. The theory of relative canonical helicity transport (You 2012) accounts for gauge dependence and generalizes the concepts of helicity evolution to open driven configurations. The theory demonstrates that the species' canonical helicities are coupled, that a system's total canonical helicity $\mathbb{K} \equiv \sum_\sigma K_\sigma$ is the appropriate system invariant, and that canonical helicity can be transferred between species.

Gauge-invariant canonical helicity is defined in a manner analogous to relative magnetic helicity (Finn & Antonsen 1985) with

$$K_{\sigma rel} \equiv \oiiint (\mathbf{P}_\sigma - \mathbf{P}_{\sigma \text{ref}}) \cdot (\mathbf{\Omega}_\sigma + \mathbf{\Omega}_{\sigma \text{ref}}) dV \tag{6}$$

where $\mathbf{P}_\sigma$ and $\mathbf{\Omega}_\sigma$ are the canonical momentum and canonical vorticity, respectively, while $\mathbf{P}_{\sigma ref}$ and $\mathbf{\Omega}_{\sigma ref}$ are reference fields to be defined later. A small difference between this definition and the definition of relative magnetic helicity is the order with which the reference fields are added or subtracted. This difference arises from the fact that canonical helicity includes physical constants while magnetic helicity does not, and that the species enthalpy is naturally useful only as a relative quantity ($h_{\sigma-} \equiv h_\sigma - h_{\sigma ref}$); thus $\mathbf{P}_{\sigma-} \equiv \mathbf{P}_\sigma - \mathbf{P}_{\sigma ref}$ must be used as opposed to $\mathbf{P}_{\sigma+} \equiv \mathbf{P}_\sigma + \mathbf{P}_{\sigma ref}$ even though both formulations are gauge invariant. To accommodate this change in sign, the reference fields must satisfy the conditions

$$\begin{cases} \nabla \times \mathbf{\Omega}_{\sigma \text{ref}}|_V = 0 \\ \mathbf{\Omega}_{\sigma \text{ref}} \cdot \hat{\mathbf{n}}|_s = -\mathbf{\Omega}_\sigma \cdot \hat{\mathbf{n}}|_s \\ \nabla \times \mathbf{P}_{\sigma \text{ref}} = \mathbf{\Omega}_{\sigma \text{ref}} \end{cases} \tag{7}$$

i.e. the normal component of the reference potential field $\mathbf{\Omega}_{\sigma ref}$ must be opposite to that of $\mathbf{\Omega}_\sigma$ at the boundaries. A convenient expansion of equation (6) shows that relative canonical helicity can be thought of as the weighted sum of three helicity components

$$K_{\sigma\text{rel}} = m_\sigma^2 \mathcal{H}_{\sigma\text{rel}} + m_\sigma q_\sigma \mathcal{X}_{\sigma\text{rel}} + q_\sigma^2 \mathcal{K}_{\text{rel}} \tag{8}$$

where $\mathcal{H}_{\sigma\text{rel}} \equiv \oiiint \mathbf{u}_{\sigma-} \cdot \mathbf{\omega}_{\sigma+} dV$ is the fluid relative kinetic helicity, $\mathcal{X}_{\sigma\text{rel}} \equiv \oiiint (\mathbf{u}_{\sigma-} \cdot \mathbf{B}_+ + \mathbf{u}_{\sigma+} \cdot \mathbf{B}_-) dV$ is the relative cross-helicity and $\mathcal{K}_{\text{rel}} \equiv \oiiint \mathbf{A}_- \cdot \mathbf{B}_+ dV$ is the relative magnetic helicity. For the case of massless electrons in a reduced two-fluid plasma, the electron canonical helicity is simply magnetic helicity weighted by the electrical charge $e^2$.

The evolution of relative canonical helicity in a given volume is obtained by taking the derivative of equation (6) and using equations (3) and (4) giving

$$\begin{aligned} \frac{dK_{\sigma\text{rel}}}{dt} = & -\oiiint (\mathbf{\Sigma}_{\sigma+} \cdot \mathbf{\Omega}_{\sigma-} + \mathbf{\Sigma}_{\sigma-} \cdot \mathbf{\Omega}_{\sigma+}) dV - \oiint h_{\sigma-} \mathbf{\Omega}_{\sigma+} \cdot d\mathbf{S} \\ & - \oiint \mathbf{P}_{\sigma-} \times \frac{\partial \mathbf{P}_{\sigma+}}{\partial t} \cdot d\mathbf{S} + \oiint (\mathbf{P}_{\sigma-} \cdot \mathbf{\Omega}_{\sigma+}) \mathbf{u}_\sigma \cdot d\mathbf{S} \end{aligned} \tag{9}$$

Each integral on the right-hand side represents a dissipative term, a battery term, an inductive term, and a Leibniz term due to the motion $\boldsymbol{u}_\sigma$ of the boundary of the system, respectively. Here, we consider helicity injection through the ends of a canonical vorticity flux tube, ignoring dissipation, induction, and motion of the boundary, leaving only the battery term. Equation (9) can therefore be written explicitly to distinguish between the flow vorticity tube component and the magnetic flux tube component

$$\dot{K}_{\sigma\text{rel}} = m_\sigma \Delta h \mathcal{F}_\sigma + q_\sigma \Delta h \psi \tag{10}$$

assuming a uniform enthalpy over the tube cross-sectional area and $\Delta h$ is the enthalpy difference between the ends of the flux tube. The ratio between the two terms on the right-hand side of equation (10) can be defined as a fractional canonical helicity injection threshold (You 2012)

$$\bar{K}_{\text{thr}} \equiv \frac{|m_\sigma \Delta h \mathcal{F}_\sigma|}{|q_\sigma \Delta h \psi|} \sim \frac{m_\sigma u_\sigma}{q_\sigma L B} \sim \frac{\rho_{L\sigma}}{L} \sim \frac{1}{S^*} \tag{11}$$

where the size parameter $S^* = L/\rho_{L\sigma}$ represents the ratio of the system scale length $L$ to the species' Larmor radius $\rho_{L\sigma}$. For large size parameter $S^*$, a given enthalpy difference $\Delta h$ applied to a canonical flux tube will channel canonical helicity preferentially into the magnetic flux tube component $\psi$, and for a small size parameter $S^*$, the same enthalpy difference $\Delta h$ will channel canonical helicity into the flow component $\mathcal{F}_\sigma$. This result provides a first-principles explanation for why the bifurcation in compact-torus merging experiments depends on the size parameter (Kawamori 2005).

From the point of view of a reduced two-fluid plasma ($m_e \to 0$), this transport model shows magnetic helicity (now topologically equivalent to electron canonical helicity) can be transferred to ion kinetic helicity while preserving the total relative canonical helicity of the system. Physically, this topological argument would be manifested as an unwinding of magnetic fields driving helical flows, and vice versa, as a winding of magnetic fields driving axial flows. Conversion between one helicity and the other would depend on the scale lengths (size parameter) where the enthalpy is applied. In our flared plasma jet geometry, the applied enthalpy is the electrostatic potential or kinetic energy, manifested as flows and electrical current density in the jet, so the profile of the current density should play an important role in driving kinetic-, cross- or magnetic helicity.

**Table 1**
Poloidal Flux Functions and Flux Coordinates for Pertinent Vector Fields

| Flux Coordinates | Vector Field General | Magnetic | Fluid (Incompressible) | Fluid Vorticity | Canonical Vorticity |
|---|---|---|---|---|---|
| $\hat{e}_1$ | $\nabla f / |\nabla f|$ | $\hat{\psi}$ | $\hat{\psi}_\sigma$ | $\hat{\mathcal{F}}_\sigma$ | $\hat{\Psi}_\sigma$ |
| $\hat{e}_2$ | $\hat{\theta} = r\nabla\theta$ | $\hat{\theta}$ | $\hat{\theta}$ | $\hat{\theta}$ | $\hat{\theta}$ |
| $\hat{e}_3$ | $\dfrac{\nabla f \times \nabla\theta}{|\nabla f \times \nabla\theta|}$ | $\hat{b}_{pol}$ | $\hat{u}_{\sigma pol}$ | $\hat{\omega}_{\sigma pol}$ | $\hat{\Omega}_{\sigma pol}$ |
| Poloidal flux function | $f = \oiint \boldsymbol{X} \cdot d\boldsymbol{S}$ | $\boldsymbol{X} = \boldsymbol{B}$ | $\boldsymbol{X} = \boldsymbol{u}_\sigma$ | $\boldsymbol{X} = \boldsymbol{\omega}_\sigma$ | $\boldsymbol{X} = \boldsymbol{\Omega}_\sigma$ |

## 3. VISUALIZATION OF CANONICAL FLUX TUBES

As opposed to magnetic flux tubes, the concept of canonical flux tubes is a little more abstract, so we develop a numerical visualization tool to examine the shapes of tubes and canonical fields for a flared magnetic flux tube carrying various profiles of flow and current density. The tool also calculates the relative canonical helicity content of the jet to examine the individual contribution from each helicity component to the total canonical helicity. This quasi-static visualization code is not designed to elucidate the full, self-consistent dynamics of canonical flux tubes, but to reveal the corresponding canonical viewpoint for a driven magnetic flux tube as current or fluid momentum is ramped up. This is similar to how the evolution of magnetic flux tubes in force-free plasmas are traced as current is ramped up (Jarboe et al. 2006), but with the important distinction that the canonical (and magnetic) fields used here are not force free. For example, the evolution of a stable screw pinch or a force-free state can be approximated by sweeping $\lambda = \mu_0 I/\psi$ values in the equation $\nabla \times \boldsymbol{B} = \lambda \boldsymbol{B}$ ($I$ is the total current carried by the configuration, $\mu_0$ is the permeability of free space, and $\psi$ is the magnetic flux in the configuration) whereas a general magnetic flux tube can kinematically evolve through various non-force-free states as axial current is applied. It is in the latter manner that we vary the enthalpy boundary conditions (i.e. axial current profile and fluid momentum) to investigate the geometry and topology of canonical fields with respect to magnetic and fluid velocity fields.

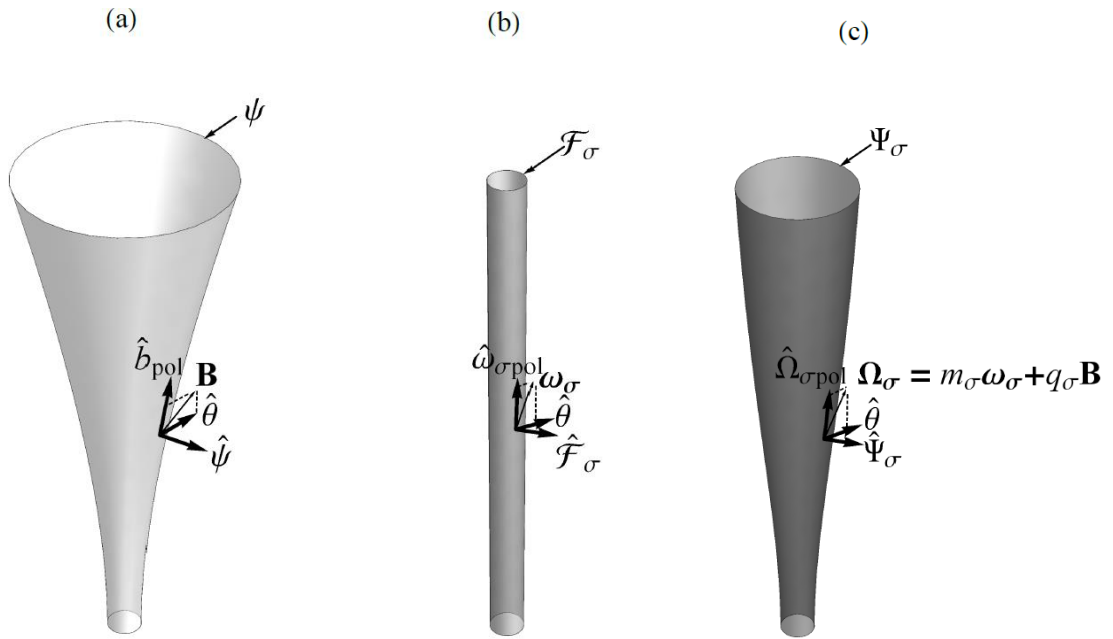

**Figure 1.** Example flux coordinate directions for (a) a magnetic flux tube $\psi$, (b) a flow vorticity flux tube $\mathcal{F}_\sigma$ of species $\sigma$, and (c) canonical vorticity flux tube $\Psi_\sigma$ for the species $\sigma$. The directions are determined by the scalar flux functions (Table 1). The canonical vorticity flux coordinates in panel (c) are the weighted sum of the magnetic and flow vorticity components of panels (a) and (b).

For simplicity, the plasma jets are assumed to be axisymmetric. Any solenoidal vector field can therefore be decomposed into poloidal ($\hat{r}, \hat{z}$) and toroidal ($\hat{\theta}$) scalar flux functions ($f$ and $g$ respectively) such that a given vector field $\boldsymbol{X}$ can be written as

$$\boldsymbol{X} = \boldsymbol{\nabla} f \times \frac{\hat{\boldsymbol{\theta}}}{r} + g \frac{\hat{\boldsymbol{\theta}}}{r}. \tag{12}$$

Table 1 and Figure 1 illustrate how a right-handed orthogonal coordinate system can be constructed based on the poloidal flux function $f$ for a given vector field and applied to four of our system's vector fields (magnetic field, fluid flow field, fluid vorticity field, canonical vorticity field).

Furthermore, we assume that our plasma jet has a flared magnetic field and carries magnetic flux in a tube of finite length and no external flows. There are thus three possible free parameters for which we choose the following (You 2014). The first free parameter is the poloidal (vacuum) magnetic field. To provide a reasonably realistic flaring shape, the poloidal magnetic field is determined by a hypothetical current loop located at $z = 0$ and oriented parallel to the z-axis such that the magnetic vector potential is given by the single-turn current loop solution

$$A_\theta(r,z) = \frac{\mu_0 I_{vac} a_{vac}}{\pi \sqrt{a_{vac}^2 + r^2 + z^2 + 2r a_{vac}}} \left( \frac{(2-k^2)K(k^2) - 2E(k^2)}{k^2} \right) \tag{13}$$

where $k^2 = 4 r a_{vac}/(a_{vac}^2 + r^2 + z^2 + 2r a_{vac})$, $I_{vac}$ and $a_{vac}$ are the ring current and radius, respectively, and $K(k^2)$ and $E(k^2)$ are the complete elliptic integrals of the first and second kind. Because of azimuthal symmetry, equation (13) can be used to obtain the poloidal magnetic flux function $\psi = 2\pi r A_\theta$ and the poloidal magnetic field

$$\boldsymbol{B_{pol}} = \frac{1}{2\pi} \left( \boldsymbol{\nabla} \psi \times \frac{\hat{\boldsymbol{\theta}}}{r} \right). \tag{14}$$

The jet is radially bounded by a magnetic flux surface $\psi_{jet}$, where $0 < \psi_{jet} < \psi_{loop}$ and $\psi_{loop}$ is the total magnetic flux through the hypothetical current loop. This defines a magnetic flux tube with an axially varying radius $a(z)$, i.e. the flare. The jet length is defined as $l$ and chosen to be several times the jet radius.

The second free parameter of the system is the current density $\boldsymbol{j}$. Since the system is fixed with no time-changing magnetic flux, we can suppose that $j_\theta$ is zero and the current is purely poloidal. The shape of the current profile can be defined arbitrarily and its effect on the canonical fields is a principal focus of this paper. Assuming the current flows along the magnetic flux surface, i.e. $\boldsymbol{j_{pol}}(\psi)$, and $\boldsymbol{\nabla} \cdot \boldsymbol{j_{pol}} = 0$ we assume that the profile is self-similar in the axial direction and introduce a normalized radial coordinate $\eta(r,z) = r/a(z)$ to build a function that describes the radial current profile $j_z(\eta)$. We examine four profiles: a uniform current profile, a diffuse core current profile, a diffuse-core-with-skin current, and a skin-only current profile (Figure 2). Defining a total jet current $I_{jet}$, the axial current density as a function of radial and axial position can be solved from

$$\begin{aligned} I_{jet} &= 2\pi \int_0^{a(z)} j_z(\eta) j_0(z) r dr \\ &= 2\pi \int_0^{a(z)} j_z(r,z) r dr \end{aligned} \tag{15}$$

where $j_0(z)$ is a scaling factor that depends on $I_{jet}$ and the axial position. From here, the radial current density, $j_r$ can be inferred and the toroidal magnetic field calculated with Ampère's law

$$B_\theta(r,z) = \frac{\mu_0 I(r,z)}{2\pi r}. \tag{16}$$

The final free parameter is the magnitude of the electron flow velocity. Because we are considering a reduced two-fluid where electrons are massless, electrons are tied to the magnetic field lines (i.e. $\boldsymbol{u_e} = u_e\hat{\boldsymbol{b}}$, where $\hat{\boldsymbol{b}} = \boldsymbol{B}/|\boldsymbol{B}|$) with a velocity determined, in principle, by some parallel electric field or pressure gradients in equation (3). In Section 4 and 5, we will choose a constant as a first approximation. Assuming a uniform density plasma in which $n_i = n_e = n_0$, the current density and electron flow velocity then implicitly define the ion fluid velocity as

$$\boldsymbol{u_i} = \frac{1}{Z}\left(\frac{\boldsymbol{j}}{n_0 e} + \boldsymbol{u_e}\right) \tag{17}$$

where $Z$ is the degree of ionization. Flow vorticities are obtained by taking the circulation of the velocity fields $\boldsymbol{\omega_\sigma} = \nabla \times \boldsymbol{u_\sigma}$, and the poloidal component of the magnetic vector potential $\boldsymbol{A_{pol}} = \mu_0/(4\pi) \oiiint (\boldsymbol{j}\, d^3r'/|\boldsymbol{r}-\boldsymbol{r}'|)$ is found by integrating the current density over the jet volume.

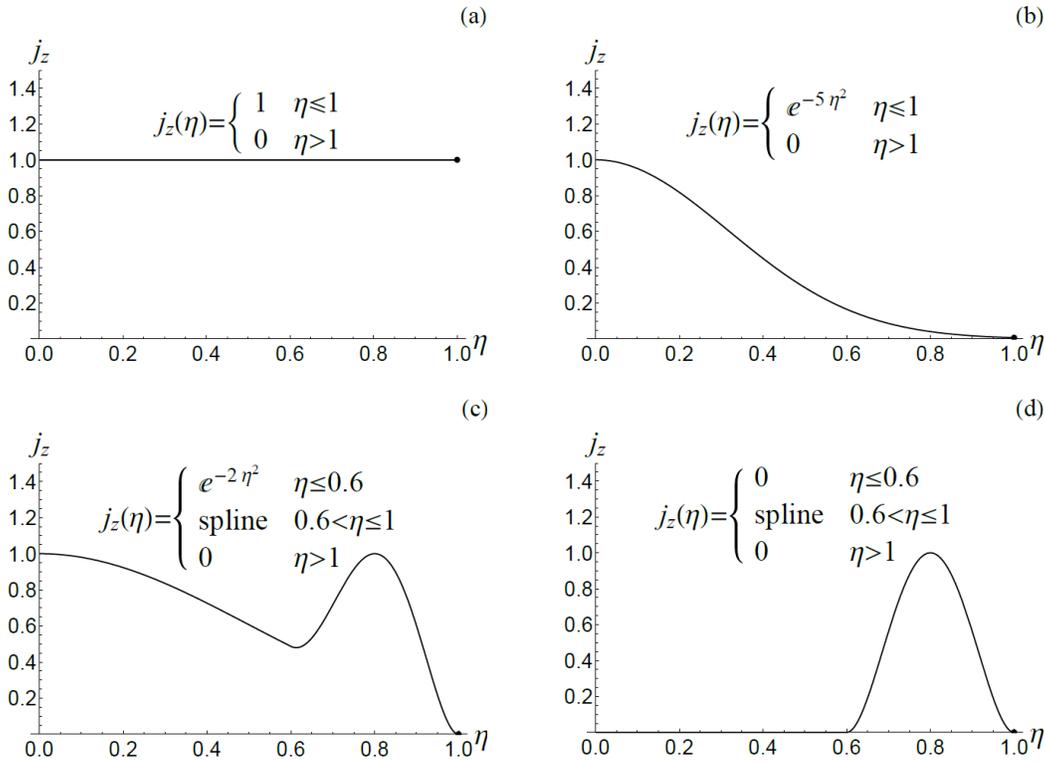

**Figure 2.** Shape of the axial current density $j_z$ as a function of normalized radius $\eta$ for (a) the uniform current profile, (b) the diffuse core current profile, (c) the diffuse-core-with-skin, and (d) skin-only current profile. The splines used in (c) and (d) are fifth-order interpolating polynomials set to match the value and slope of the core current at $\eta = 0.6$ with a maximum at $\eta = 0.8$ and going smoothly to zero at $\eta = 1$. These profiles are scaled by a multiplicative constant $j_0(z)$ to retrieve the total jet current $I_{\text{jet}}$ when integrated over the cross-sectional area at any axial location (Equation (15)).

With these field components defined, the canonical fields can be constructed for the electrons and the ions as follows.

$$\begin{cases} \boldsymbol{P}_e = -e\boldsymbol{A} \\ \boldsymbol{\Omega}_e = -e\boldsymbol{B} \end{cases} \quad \begin{cases} \boldsymbol{P}_i = m_i \boldsymbol{u}_i + Ze\boldsymbol{A} \\ \boldsymbol{\Omega}_i = m_i \boldsymbol{\omega}_i + Ze\boldsymbol{B}. \end{cases} \tag{18}$$

These vector fields are converted from cylindrical to Cartesian coordinates, and their streamlines are plotted parametrically by solving the equation

$$\frac{\partial \boldsymbol{r}(s)}{\partial s} = \boldsymbol{X}(\boldsymbol{r}(s)) \tag{19}$$

where $s$ is a parametric index, $\boldsymbol{r}(s)$ is the position vector, and $\boldsymbol{X}$ is the vector field. Canonical vorticity flux surfaces are calculated in the same manner as the magnetic flux surface by setting the canonical vorticity flux to be a constant, in this case $\Psi_\sigma = 2\pi a_0 P_\theta(a_0, 0)$, i.e. the total canonical vorticity flux through the jet radius at $z = 0$. For massless electrons $\Psi_e = -e\psi$ and the electron canonical vorticity flux surfaces are indistinguishable from magnetic flux surfaces. For the ions, however, finite momentum can lead to significant departures of the ion canonical vorticity flux surfaces from the magnetic flux surfaces.

For flexibility, we normalize our quantities and the governing equations by letting $\overline{\boldsymbol{\nabla}} = l_0 \boldsymbol{\nabla}$, $\overline{\boldsymbol{B}} = \boldsymbol{B}/B_0$, $\bar{q}_\sigma = q_\sigma/e$, $\bar{m}_\sigma = m_\sigma/m_i$, $\overline{\boldsymbol{A}} = \boldsymbol{A}/(B_0 l_0)$, $\overline{\boldsymbol{u}}_\sigma = \boldsymbol{u}_\sigma/v_a$, $\bar{t} = t v_a/l_0$, $\bar{\phi} = \phi e/(m_i v_a^2)$, $\bar{p}_\sigma = 2\mu_0 p_\sigma/B_0^2$, where $v_a = B_0/\sqrt{\mu_0 n_0 m_i}$. Here $l_0$ is a convenient system scale length and $B_0$ is a convenient scale magnetic field strength. Neglecting friction, the normalized canonical equation of motion (3) becomes

$$\frac{\partial \overline{\boldsymbol{P}}_\sigma}{\partial \bar{t}} - \overline{\boldsymbol{u}}_\sigma \times \overline{\boldsymbol{\Omega}}_\sigma = -\overline{\boldsymbol{\nabla}} \bar{h}_\sigma \tag{20}$$

with the normalized canonical fields

$$\begin{cases} \overline{\boldsymbol{P}}_\sigma = \bar{m}_\sigma \overline{\boldsymbol{u}}_\sigma + S_0 \bar{q}_\sigma \overline{\boldsymbol{A}} \\ \overline{\boldsymbol{\Omega}}_\sigma = \bar{m}_\sigma \overline{\boldsymbol{\omega}}_\sigma + S_0 \bar{q}_\sigma \overline{\boldsymbol{B}} \\ \bar{h}_\sigma = \frac{1}{2}\bar{m}_\sigma \bar{u}_\sigma^2 + \bar{q}_\sigma \beta_{ES} + \frac{1}{2}\beta_\sigma \end{cases} \tag{21}$$

The parameter $S_0 = eB_0 l_0/(m_i v_a) = l_0/r_{L0}$ is a size parameter for an Alfvénic ion travelling in the characteristic magnetic field $B_0$; the parameter $\beta_{ES} = e\phi/m_i v_a^2$ is an electrostatic beta expressing the relative strength of the electrostatic field to the kinetic energy; and the parameter $\beta_\sigma = 2\mu_0 p_\sigma/B_0^2$ is the usual species beta expressing the relative strength of fluid pressure to magnetic pressure. Eliminating the magnetic field in equation (21) gives the ion canonical vorticity as

$$\overline{\boldsymbol{\Omega}}_i = \overline{\boldsymbol{\omega}}_i + \bar{m}_e \overline{\boldsymbol{\omega}}_e - \overline{\boldsymbol{\Omega}}_e \tag{22}$$

which shows how canonical quantities for each species are coupled to each other through the magnetic field, i.e. the ion canonical vorticity is antiparallel to the electron canonical vorticity with correction factors based on the strength of the species' flow vorticities.

The dimensionless relative canonical helicity is defined from the normalized canonical fields as

$$\overline{K}_{\sigma\mathrm{rel}} \equiv \oiiint \overline{\boldsymbol{P}}_{\sigma-} \cdot \overline{\boldsymbol{\Omega}}_{\sigma+} \, d\overline{V} \tag{23}$$

where the normal component of the reference potential field $\overline{\boldsymbol{\Omega}}_{\sigma ref}$ must be opposite that of $\overline{\boldsymbol{\Omega}}_\sigma$ on the normalized system boundary (equation 7). Assigning the integration volume to be the driven magnetic flux tube, this boundary condition can be broken up into magnetic and flow vorticity components. This allows for the calculation of normalized helicity components

$$\overline{K}_{\sigma rel} \equiv \overline{m}_\sigma^2 \overline{\mathcal{H}}_{\sigma rel} + \overline{m}_\sigma \overline{q}_\sigma S_0 \overline{\mathcal{X}}_{\sigma rel} + \overline{q}_\sigma^2 S_0^2 \overline{\mathcal{K}}_{rel} \tag{24}$$

as outlined in equation (8). Because the magnetic field lines are tangential to the side boundary of the integration volume, the only conditions on the reference magnetic field are

$$\begin{cases} \nabla \times \overline{\boldsymbol{B}}_{\mathbf{ref}}|_V = 0 \\ \overline{\boldsymbol{B}}_{\mathbf{ref}} \cdot \hat{\boldsymbol{z}}|_{\overline{l}} = -\overline{\boldsymbol{B}} \cdot \hat{\boldsymbol{z}}|_{\overline{l}} \\ \overline{\boldsymbol{B}}_{\mathbf{ref}} \cdot (-\hat{\boldsymbol{z}})|_0 = -\overline{\boldsymbol{B}} \cdot (-\hat{\boldsymbol{z}})|_0. \end{cases} \tag{25}$$

This condition is satisfied by letting $\overline{\boldsymbol{B}}_{\mathbf{ref}} = -\overline{\boldsymbol{B}}_{\mathbf{pol}}$ since the poloidal magnetic field is a potential (vacuum) field that satisfies these boundary conditions. The reference magnetic vector potential is then $\overline{\boldsymbol{A}}_{\mathbf{ref}} = -\bar{A}_\theta \hat{\boldsymbol{\theta}}$. The flow vorticity field on the other hand is not always tangential to the side surface of the magnetic flux tube, so the correct reference field must be solved from Laplace's equation

$$\begin{cases} \nabla^2 \phi_\sigma = 0 \\ \nabla \phi_\sigma \cdot \hat{\boldsymbol{n}}|_s = -\overline{\boldsymbol{\omega}}_{\sigma ref} \cdot \hat{\boldsymbol{n}}|_s \end{cases} \tag{26}$$

where $\nabla \phi_\sigma = \overline{\boldsymbol{\omega}}_{\sigma ref}$ and $\hat{\boldsymbol{n}}$ is the surface normal. The solution to equation (26) is not unique up to a constant, but because we are only interested in the gradient of $\phi_\sigma$ it is sufficient for defining our reference field. The corresponding flow reference field is obtained from axisymmetry using $\overline{\boldsymbol{u}}_{\sigma ref} = \bar{f}_{\sigma ref}/2\pi\bar{r}\ \hat{\boldsymbol{\theta}}$, where $\bar{f}_{\sigma ref}$ is the reference flow vorticity flux. The reference canonical fields can then be constructed from the weighted combination of the magnetic and flow reference field components.

　　Figure 3 shows, for the case of a jet with a uniform radial current profile, the five equivalent viewpoints of the concept of a flux tube in a magnetized flowing plasma. The current-carrying magnetic flux tube $\psi \equiv \int \boldsymbol{B} \cdot d\boldsymbol{S}$ (Figure 3(a)) is flared with twisted magnetic field lines due to the superposition of a poloidal current with a poloidal magnetic dipole field. The center of mass velocity $\boldsymbol{U}$ (equivalent to the ion velocity in a reduced two-fluid where $m_e = 0$) is stuck to the magnetic flux surfaces but not the field lines. Figure 3(b) shows the same system as a pair of electron and ion flow tubes $\psi_\sigma \equiv \int \boldsymbol{u}_\sigma \cdot d\boldsymbol{S}$. The electron flow tube is geometrically coincident ($\Leftrightarrow$) with the magnetic flux tube since the electrons are here taken to be massless and thus flow along magnetic field lines at a constant velocity. The ions must compensate for the electron direction to carry the poloidal current, so the helical ion flow streamlines have a steeper pitch angle $\theta = \tan^{-1}(\bar{u}_{iz}/\bar{u}_{i\theta})$ than the helical electron flow streamlines. Since the ion flow streamlines also remain on the magnetic flux surface, the ion and electron flow tubes both coincide with the magnetic flux tube. Figure 3(c) shows the same system as a pair of electron and ion flow vorticity flux tubes $\mathcal{F}_\sigma \equiv \int \boldsymbol{\omega}_\sigma \cdot d\boldsymbol{S}$. Because $j_\theta = 0$, both species have the same radial gradient in toroidal velocity and thus share identical axial flow vorticity $\omega_{\sigma z} = r^{-1}\partial/\partial r(ru_{\sigma\theta})$. At moderate to large jet currents, $\omega_{\sigma z}$ dominates the poloidal fluid vorticity and ion and electron vorticity flux tubes are nearly identical. Since the jet is flared, a toroidal fluid vorticity $\omega_{\sigma\theta} = \partial u_{\sigma r}/\partial z - \partial u_{\sigma z}/\partial r$ gives rise to helical streamlines that have slightly different pitch angles for each species. Figure 3(d) shows the same system as a pair of canonical momentum flux tubes $\mathcal{P}_\sigma \equiv \int \boldsymbol{P}_\sigma \cdot d\boldsymbol{S}$. The streamlines are again helical with a pitch angle that increases in the axial direction as the toroidal magnetic vector potential $A_\theta$ decreases far from the axis. The flux surfaces are distinct for each species since a finite ion fluid momentum pulls ion canonical momentum flux surfaces closer to magnetic flux surfaces and keeps streamlines more twisted further up the tube (provided the ion flow is helical). Figure 3(e) shows the same system as a pair of canonical vorticity flux tubes $\Psi_\sigma \equiv \int \boldsymbol{\Omega}_\sigma \cdot d\boldsymbol{S}$. The electron canonical vorticity is antiparallel to the magnetic field and therefore geometrically coincident to the magnetic and flow velocity tubes. In contrast, the ion canonical vorticity flux tube departs from the magnetic flux surface due to the added contribution of the poloidal ion flow vorticity. All of these equivalent viewpoints range from the familiar current-carrying magnetic flux tubes to the more abstract pair of canonical flux tubes.

The strength of the canonical flux tube viewpoint (Figure 3(e)) compared to the current-carrying magnetic flux tube viewpoint (Figure 3(a)) is the ability to visualize behavior more general than permitted by MHD (You 2014) while retaining our intuition of changing geometry: for example, laboratory observations show that reconnecting magnetic flux tubes can generate new magnetic flux tubes and generate strong flows due to two-fluid or kinetic effects (Fiksel et al. 2009). Magnetic helicity conservation is based on MHD, and thus can only predict the topological evolution of magnetic flux tubes into other magnetic flux tubes but not into or from flows. On the other hand, canonical helicity conservation (Section 2) can predict the topological evolution of both magnetic and flow fields as linkage between magnetic and flow flux tubes (Steinhauer & Ishida 1997, 1998; Steinhauer et al. 2001; You 2012, 2014). Since enthalpy gradients at the boundaries can increase magnetic twist or generate helical flows depending on the uniformity of the enthalpy gradient (Equations (9) and (10)), the next section investigates the kinematic evolution of canonical flux tubes with four different enthalpy conditions at the boundaries for the case of a flared, cylindrically symmetric geometry with massless electrons.

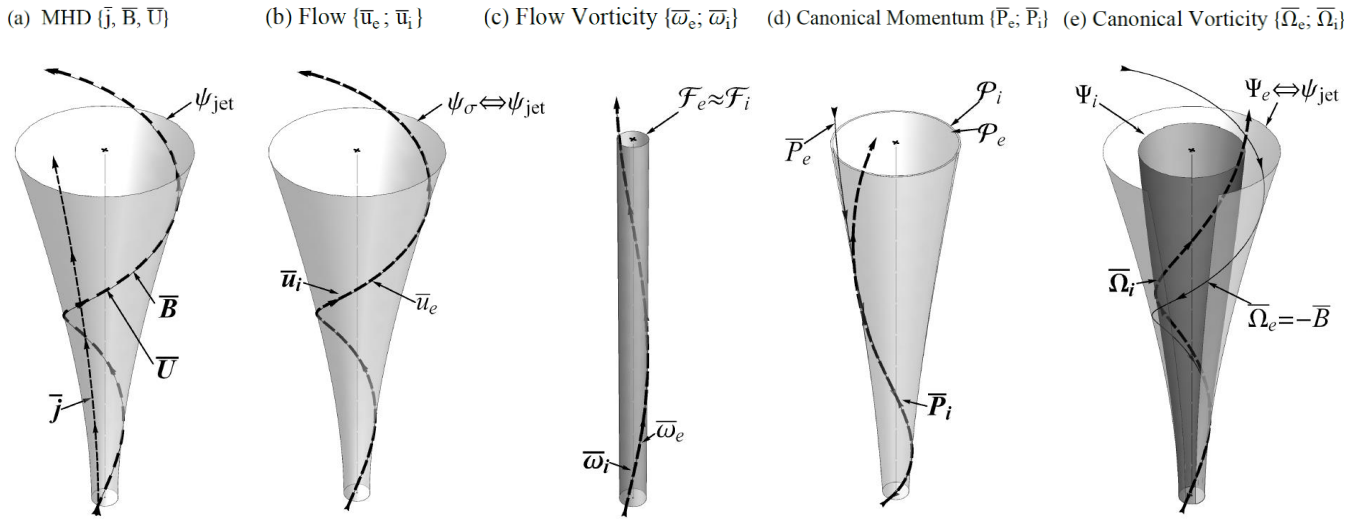

**Figure 3.** Five equivalent interpretations of flux tubes from the most familiar to the most abstract: (a) a current-carrying magnetic flux tube $\psi_{jet}$ with uniform current profile and bulk flow velocity $\overline{U}$ for $l_0^{-1}\lambda \approx 2.5$ and $|\overline{u}_e| = 3$; (b) ion and electron fluid velocity flux tubes $\psi_\sigma$ with flows restricted to magnetic flux surfaces; (c) ion and electron fluid vorticity flux tubes $\mathcal{F}_\sigma$; (d) ion and electron canonical momentum flux tubes $\mathcal{P}_\sigma$; (e) ion and electron canonical vorticity flux tubes $\Psi_\sigma$. Here, the electron flux tube is coincident with the magnetic flux tube (light gray in panels a, b, and e) because of negligible electron mass while the ion flux tube of interest is highlighted (dark gray in panel e).

## 4. KINEMATIC EVOLUTION OF FLARED PLASMA JET

The enthalpy driving the evolution of the magnetic flux tube (Figure 3(a)) and the canonical flux tube (Figure 3(e)) is chosen to be the plasma jet current expressed as the dimensionless current parameter $\bar{\lambda} = \mu_0 I_{\text{jet}}/\psi_{jet} l_0$ where $\mu_0$ is the permittivity of free space, $I_{\text{jet}}$ is the total current inside the jet, $\psi_{jet}$ is the total (fixed) magnetic flux, and $l_0$ is the characteristic scale length. The enthalpy is distributed across the cross-section of the tube as one of four possible current profiles (Figure 2) and either a sub- or super-Alfvénic electron flow velocity $|\bar{u}_e|$ (Table 2). Section 4.1 presents the cases with uniform and diffuse core currents. Section 4.2 presents the cases with diffuse-core-with-skin and skin-only currents. For all cases, the system constants are set to $m_i = m_{\text{proton}}$, $Z = 1$, $I_{\text{vac}} = 10^4$ A, $a_{\text{vac}} = 5$ m, $n_0 = 10^{20}\ m^{-3}$, $l = 15$ m, and $\psi_{\text{jet}} = 1$ mWb. The normalization values are chosen to be the magnetic field strength on axis $B_0 \sim 1.25$ mT and the jet radius at the footpoint $l_0 \sim 0.5$ m; resulting in an Alfvèn velocity of 2.74 $km\ s^{-1}$ and a size parameter of $S_0 \approx 22$. To illustrate shear between flux surfaces, streamlines are also presented on an interior magnetic ($\psi_{\text{inner}} = 0.25$ mWb) and ion canonical flux tube ($\Psi_{i,inner}$) both defined to have the same radius at the starting footpoint $z = 0$.

**Table 2**
Selected Jet Parameters for Kinematic Evolution Calculations and Corresponding Figures

| Enthalpy (Current Profile, Fig. 2) | Electron Velocity: $\|\bar{u}_e\|$ | MHD: $\{J; B; U\}$ $\bar{\lambda} = 1$ | $\bar{\lambda} = 10$ | Canonical: $\{\Omega_e; \Omega_i\}$ $\bar{\lambda} = 1$ | $\bar{\lambda} = 10$ |
|---|---|---|---|---|---|
| Uniform | $\|\bar{u}_e\| = 2/3$ | Fig. 4a | Fig. 4b | Fig. 5a | Fig. 5b |
|  | $\|\bar{u}_e\| = 4$ | Fig. 4c | Fig. 4d | Fig. 5c | Fig. 5d |
| Diffuse Core | $\|\bar{u}_e\| = 2/3$ | Fig. 4e | Fig. 4f | Fig. 5e | Fig. 5f |
|  | $\|\bar{u}_e\| = 4$ | Fig. 4g | Fig. 4h | Fig. 5g | Fig. 5h |
| Diffuse Core with Skin | $\|\bar{u}_e\| = 2/3$ | Fig. 8a | Fig. 8b | Fig. 9a | Fig. 9b |
|  | $\|\bar{u}_e\| = 4$ | Fig. 8c | Fig. 8d | Fig. 9c | Fig. 9d |
| Skin | $\|\bar{u}_e\| = 2/3$ | Fig. 8e | Fig. 8f | Fig. 9e | Fig. 9f |
|  | $\|\bar{u}_e\| = 4$ | Fig. 8g | Fig. 8h | Fig. 9g | Fig. 9h |

### *4.1 . Uniform and Diffuse Core Current Profiles*

Figure 4 presents the morphology of a flared plasma jet with uniform and diffuse core current profiles from the MHD perspective of current-carrying magnetic flux tubes at low $\bar{\lambda} = 1$ and high $\bar{\lambda} = 10$ with a bulk velocity $\bar{U}$ also effected by the choice of $|\bar{u}_e|$. In both uniform and diffuse current cases, as $\bar{\lambda}$ (i.e. jet current) is increased, the magnetic field becomes helical, with a shear in the non-uniform current case, due to the superposition of increasing poloidal current with a constant poloidal magnetic field. The bulk flow velocity $\bar{U}$ (equivalent to the ion flow velocity) is tied to magnetic flux surfaces with a steeper pitch angle than the magnetic field in regions of finite current density. This pitch angle is steeper with lower electron velocities since the ions must unwind to accommodate the poloidal current density. Because $j_\theta = 0$, the magnetic flux surfaces remain unchanged as current is increased. However, from a canonical flux tube point-of-view (Figure 5), the shape of canonical flux surfaces depends on the species flow velocity as well as on the jet current profile and magnitude (Figure 6). Figures 5(a)-(d) show the equivalent systems to Figures 4(a)-(b) (uniform current profile) for a choice of $|u_e|$ that is sub-Alfvénic (Figures 5(a)-(b)) or super-Alfévnic (Figures 5(c)-(d)). Figures 5(e)-(h) show the equivalent systems to Figures 4(c)-(d) (diffuse core current profile) for the same electron flow velocities. In all cases, the electron canonical flux tubes are geometrically coincident with the magnetic flux tubes, $\Psi_e \Leftrightarrow \psi_{jet}$, because of our massless electron assumption. In all cases, the ion canonical flux tube departs from the magnetic flux surfaces $\Psi_i \Leftrightarrow \psi_{jet}$ and becomes collimated further up away from the footpoint even though the initial cross-sectional area at the footpoint is the same as the magnetic flux tube. The collimation of the ion canonical flux tube occurs closer to the footpoint at larger currents $\bar{\lambda}$ (e.g. Figure 5(b) versus Figure 5(a)) and larger electron flow velocities $|\bar{u}_e|$ (e.g. Figure 5(d) versus Figure 5(b), and Figure 6).

Remembering that for an axisymmetric system the curl of a poloidal field is a toroidal field and vice versa, the separation of canonical flux surfaces can be explained as follows. At zero jet current, the ion flow has a small but finite toroidal circulation $\omega_{i\theta}$ because the ions flow with a constant velocity along flared magnetic field lines. Because the ion flow is purely poloidal at this point, then $\bar{\boldsymbol{\omega}}_{\mathbf{ipol}} = \mathbf{0}$ and the ion canonical vorticity flux is just the charge weighted magnetic flux $\Psi_i = q_i\psi$. With a finite jet current, the ion flows become helical and $\bar{\boldsymbol{\omega}}_{\mathbf{ipol}}$ becomes finite so the poloidal component of the ion canonical vorticity is no longer parallel to the poloidal magnetic field $\bar{\boldsymbol{\Omega}}_{\mathbf{ipol}} = \bar{\boldsymbol{\omega}}_{\mathbf{ipol}} + \bar{\boldsymbol{B}}_{\mathbf{pol}}$, causing the departure from the magnetic flux surfaces of the ion canonical vorticity flux surfaces (Figures 5 and 6).

The variation of poloidal flow vorticity $\bar{\boldsymbol{\omega}}_{\mathbf{ipol}} = \bar{\omega}_{ir}\hat{\boldsymbol{r}} + \bar{\omega}_{iz}\hat{\boldsymbol{z}}$ in space, with $\bar{\lambda}$, and with $|\bar{u}_e|$ results in the collimation of the canonical flux tube (Figure 6). The ion canonical flux tube is fully collimated when the radius of the canonical vorticity flux tube $a_{\Psi_i}(z)\sim\text{constant}$. i.e. when

$$\begin{cases} \bar{\omega}_{iz} \gg |\bar{\boldsymbol{B}}_{\text{pol}}| \\ \bar{\omega}_{iz} \gg \bar{\omega}_{ir} \end{cases} \quad (27)$$

because $\bar{\boldsymbol{\Omega}}_{\mathbf{ipol}} = \bar{\boldsymbol{\omega}}_{\mathbf{ipol}} + \bar{\boldsymbol{B}}_{\mathbf{pol}}$. The ion canonical vorticity vector points in the direction of the (flared) poloidal magnetic field modified by the poloidal flow vorticity, which can cancel the flare if equation (27) is satisfied. Because the magnitude of $\omega_{iz}$ depends on the rate at which $rB_\theta$ (and thus $ru_{e\theta}$ and $ru_{i\theta}$) increases in the radial direction, the axial component of the flow vorticity is proportional to the current profile, i.e. $\omega_{iz} \propto \partial I(r,z)/\partial r$ where $I(r,z)$ is the total current at a given radial and axial location. Since the total current increases with radius for both current profiles, the axial component of the flow vorticity $\omega_{iz} = 1/r\, \partial(ru_{i\theta})/\partial r$ is always positive so always points up the tube. The radial profile $\omega_{iz}(r)$ is dictated by the shape and strength of the current profile $j_z(r)$ and the magnitude of the electron flow velocity (curl of equation (17)). The radial component of the ion flow vorticity $\omega_{ir} = -\partial u_{i\theta}/\partial z$ is a consequence of the flaring of the jet and can be positive (point outward) or negative (point inward) at any given radial and axial location based on the shape of the current profile. For the uniform current cases of Figure 5, $\partial u_{i\theta}/\partial z \sim 0$ so $\omega_{ir} \ll \omega_{iz}$ and can be neglected. The ion canonical vorticity is then dominated by the sum of the poloidal magnetic field and axial flow vorticity, $\bar{\boldsymbol{\Omega}}_{\mathbf{ipol}} \simeq \bar{\omega}_{iz}\hat{\boldsymbol{z}} + \bar{\boldsymbol{B}}_{\mathbf{pol}}$. Close to the footpoint, the ion canonical vorticity departs little from the magnetic field because $\bar{\boldsymbol{B}}_{pol}$ is large and the flaring angle of the jet $\theta = \tan^{-1}(B_r/B_z)$ is nearly 0 (i.e. $B_z \gg B_r$). Further away from the footpoint, the poloidal magnetic field is flared ($B_z \sim B_r$) but weak $|\bar{\boldsymbol{B}}_{\text{pol}}| \to 0$, so the canonical vorticity becomes more axial $\bar{\boldsymbol{\Omega}}_{\mathbf{ipol}} \sim \bar{\omega}_{iz}\hat{\boldsymbol{z}}$ resulting in a collimated canonical flux tube $\Psi_i$. The location up the axis where the flux tube is fully collimated depends on where $\omega_{iz} \gg |\boldsymbol{B}_{pol}|$ and occurs closer to the starting footpoint at larger flow momentum $|\bar{u}_e|$ and $\bar{\lambda}$ values (Figures 6(a)-(b)).

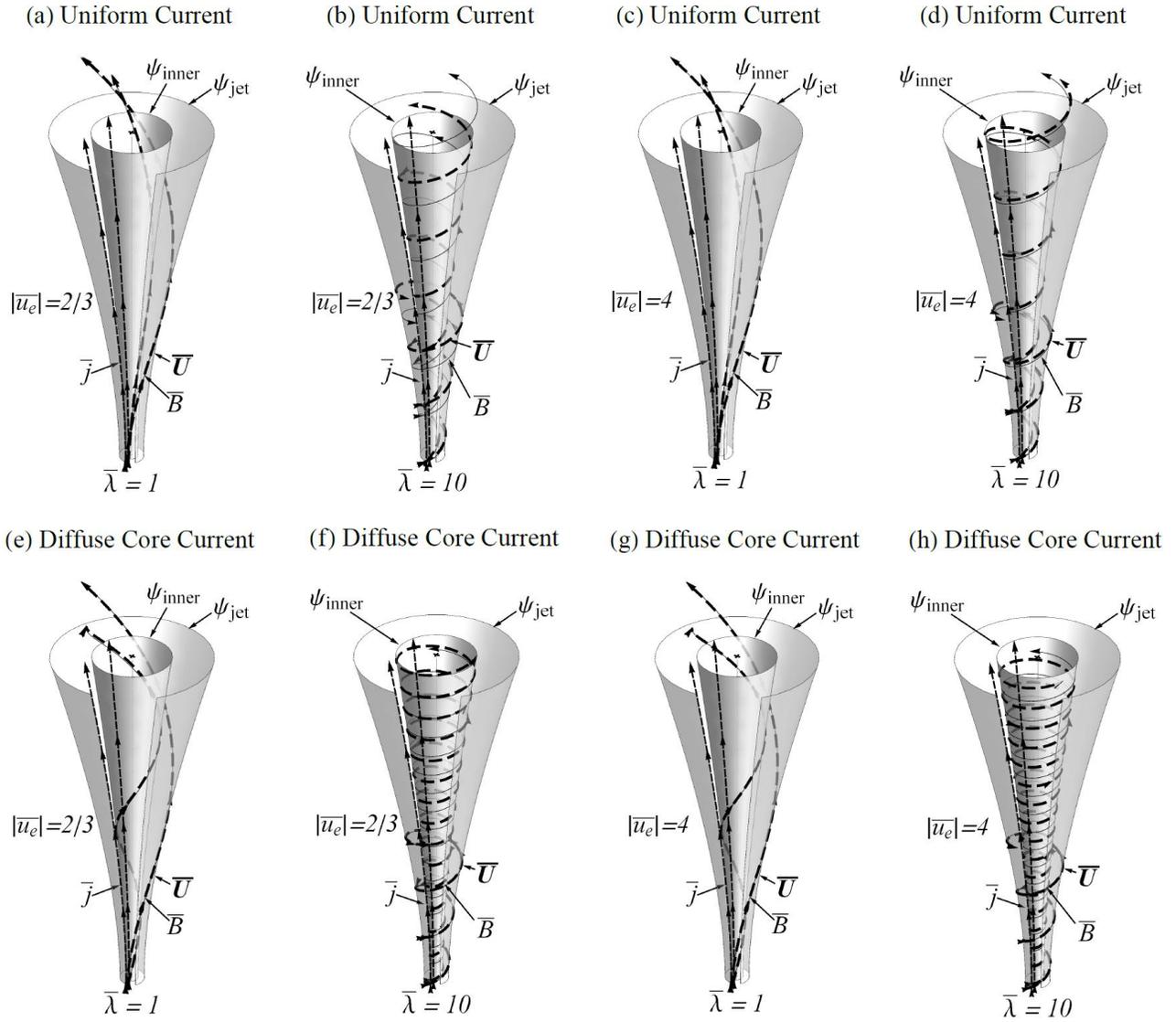

**Figure 4.** Flared magnetic flux tubes with uniform and diffuse core current profiles for sub-Alfvénic ($|\bar{u}_e| = 2/3$) and super-Alfvénic ($|\bar{u}_e| = 4$) electron flow velocities at low $\bar{\lambda} = 1$ and high $\bar{\lambda} = 10$. Thin, solid lines are magnetic field streamlines ($\bar{B}$), thin dashed lines are current density streamlines ($\bar{j}$), and bold dashed lines are bulk flow velocity streamlines ($\bar{U}$). Arrows on the streamline indicate the direction of the vector field. Streamlines on both the outer magnetic flux surface ($\psi_{jet}$) and an inner magnetic flux surface ($\psi_{inner}$) are shown; however, some streamlines on the outer magnetic flux surface have been truncated for visual clarity. Figure 5 shows the same systems from a canonical flux tube point-of-view.

Because the current density is self-similar along the magnetic flux tube, ion canonical vorticity streamlines that depart magnetic flux surfaces travel through different regions within the current profile. Different strengths of current density change the strength of the ion flow velocities (equation 17) and therefore change the contribution of $\omega_{ir}$ and $\omega_{iz}$ to the canonical vorticity $\mathbf{\Omega}_i$. For the uniform current case, $|\bar{\omega}_{ir}| \ll |\bar{\omega}_{iz}|$ everywhere and can be ignored. For the diffuse core current, however, there is a region near the exterior of the jet where $\bar{\omega}_{ir} < 0$ and $|\bar{\omega}_{ir}| \gtrsim |\bar{\omega}_{iz}|$ at low $\bar{\lambda}$ values (Figure 6(c)). The axial flow vorticity $\bar{\omega}_{iz}(r)$ is also strongly peaked on axis. These two conditions influence where ion canonical flux surfaces depart the magnetic flux surface and where complete collimation occurs. Therefore, the ion canonical flux tube collimates closer to the base of the jet in the diffuse core current case (Figure 6(c)) than for the uniform current case (Figure 6(a)) at low $\bar{\lambda}$. This is true for both sub- and super-Alfvénic electron flow velocities (Figure 5(a), (e) and Figure 5(c), (g) respectively). At high $\bar{\lambda}$ however, $\bar{\omega}_{iz}$ dominates the effect of $\bar{\omega}_{ir}$ so the collimation location and radii is not so sensitive to the current profile (Figure 5(b) versus 5(d); Figure 5(f) versus 5(e); Figure 6(b) versus 6(d)).

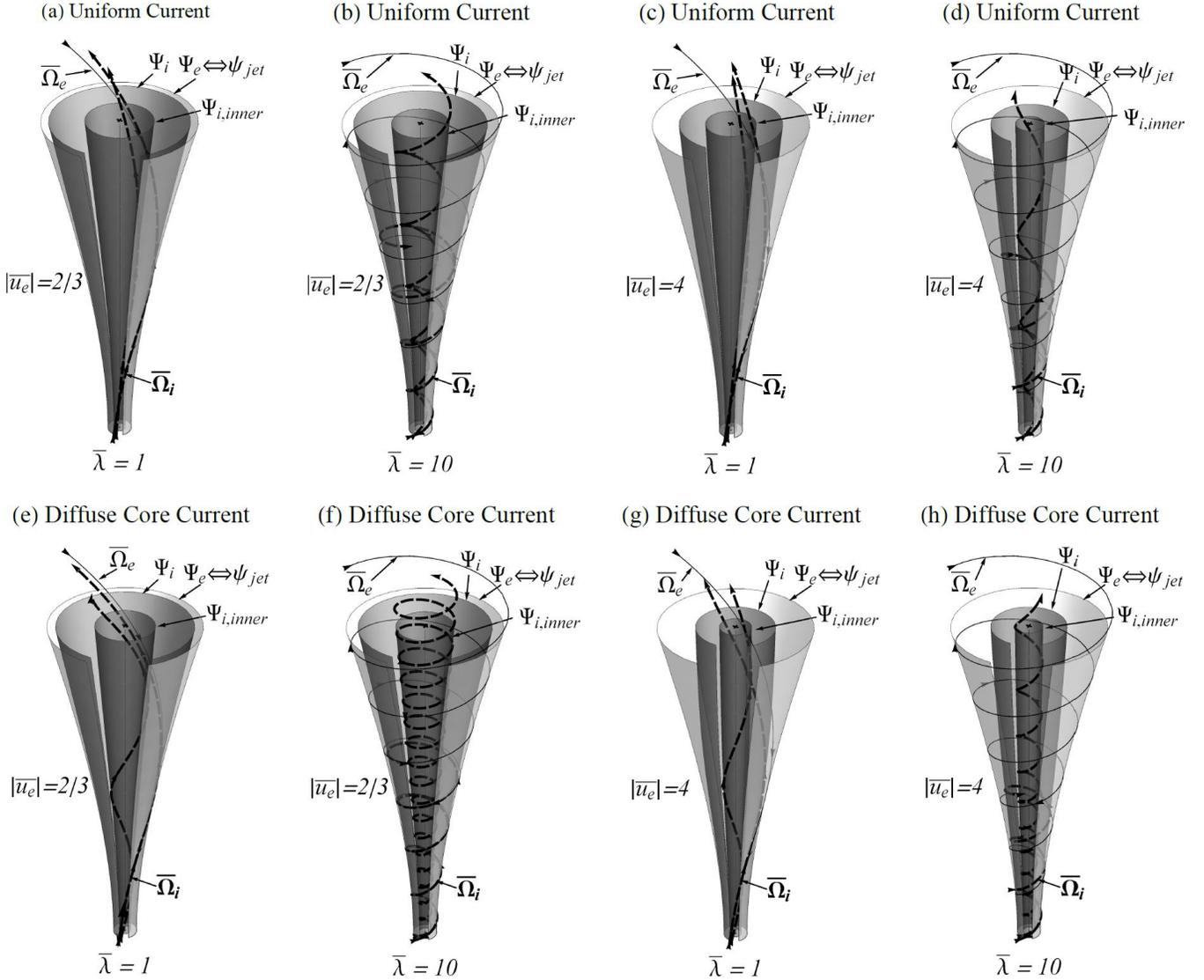

**Figure 5.** Canonical vorticity flux tubes with uniform and diffuse core current profiles for sub-Alfvénic ($|\bar{u}_e| = 2/3$) and super-Alfvénic ($|\bar{u}_e| = 4$) electron flow velocities at low $\bar{\lambda} = 1$ and high $\bar{\lambda} = 10$. These canonical flux tubes are equivalent to the corresponding magnetic flux tubes of Fig. 4. Ion canonical vorticity streamlines ($\bar{\Omega}_i$) are drawn as bold, dashed lines along two ion canonical vorticity flux tubes ($\Psi_i$ and $\Psi_{i,inner}$), while electron canonical vorticity ($\bar{\Omega}_e$) streamlines (geometrically equivalent to magnetic field streamlines), are depicted as a thin, solid line along an electron canonical flux tube ($\Psi_e \Leftrightarrow \psi_{jet}$, light gray tube). Some streamlines on the outer ion canonical vorticity flux tube ($\Psi_i$, dark gray tubes) have been truncated for visual clarity. As current is increased, ion canonical flux tubes depart from magnetic flux tube surfaces and become collimated.

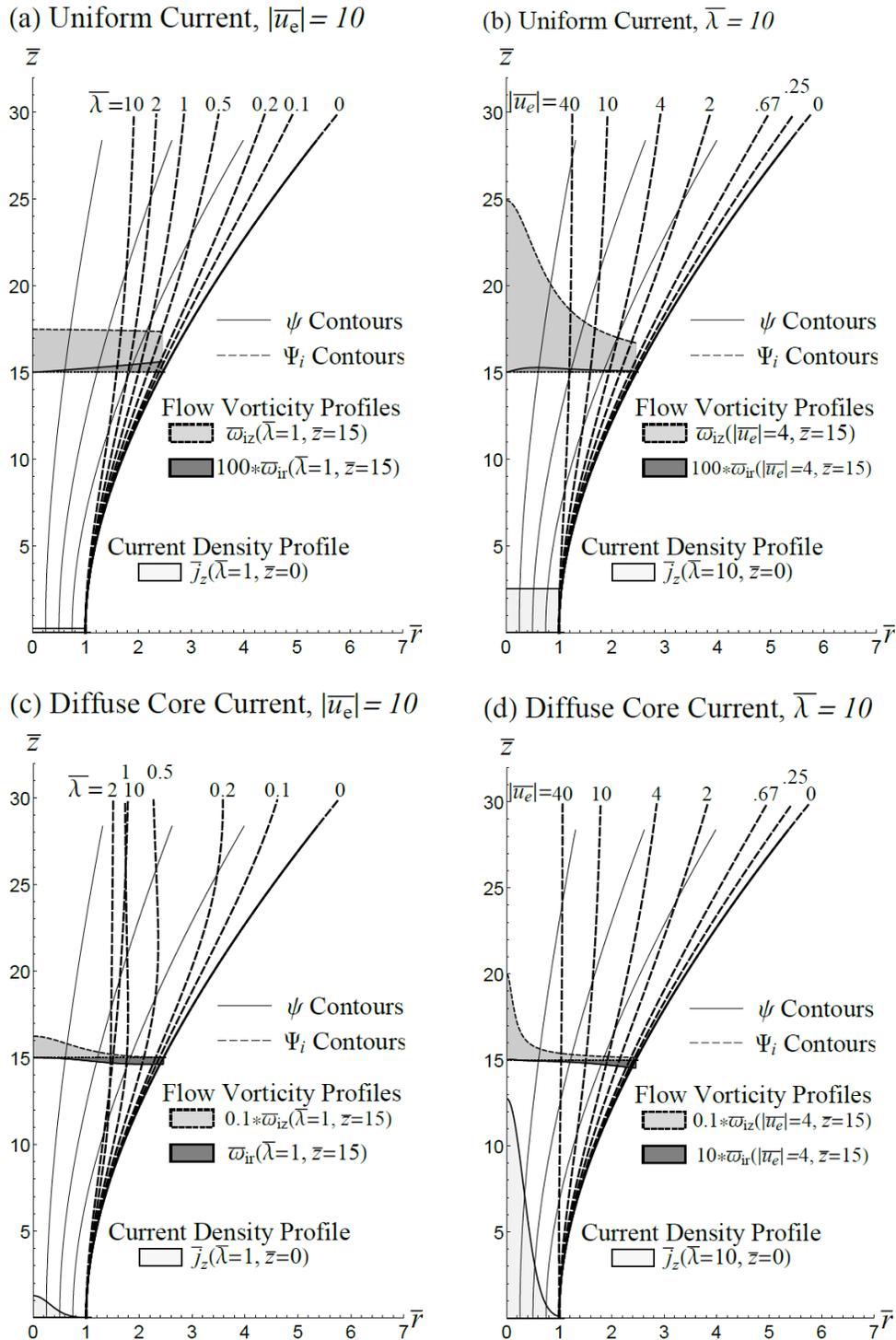

**Figure 6.** Dependence of ion canonical vorticity flux surface $\Psi_i$ (dashed lines) on total jet current $\bar{\lambda}$ (panels a, c) and on electron fluid velocity $|\bar{u}_e|$ (panels b, d) for uniform (panels a, b) and diffuse core (panels c, d) currents. For both current profiles, full collimation occurs closer to the base of the flux tube with increasing electron velocity $|\bar{u}_e|$. A prominent bulge in the canonical flux tube occurs at low $\bar{\lambda} \sim 0.5$ and high $|\bar{u}_e| = 10$ for a diffuse core current profile (panel c). Plots of current profile $\bar{j}_z(\bar{r})$ are shown at the base of the jet (lightest gray) while representative poloidal flow vorticity profiles $\bar{\omega}_{ir}(\bar{r})$ (dark gray) and $\bar{\omega}_{iz}(\bar{r})$ (light gray) are drawn at $\bar{z} = 15$. Contours of constant magnetic flux $\psi$ (solid lines) do not change with $\bar{\lambda}$ nor $|\bar{u}_e|$ and are presented to reveal passage of the canonical flux surface through different regions of the current profile.

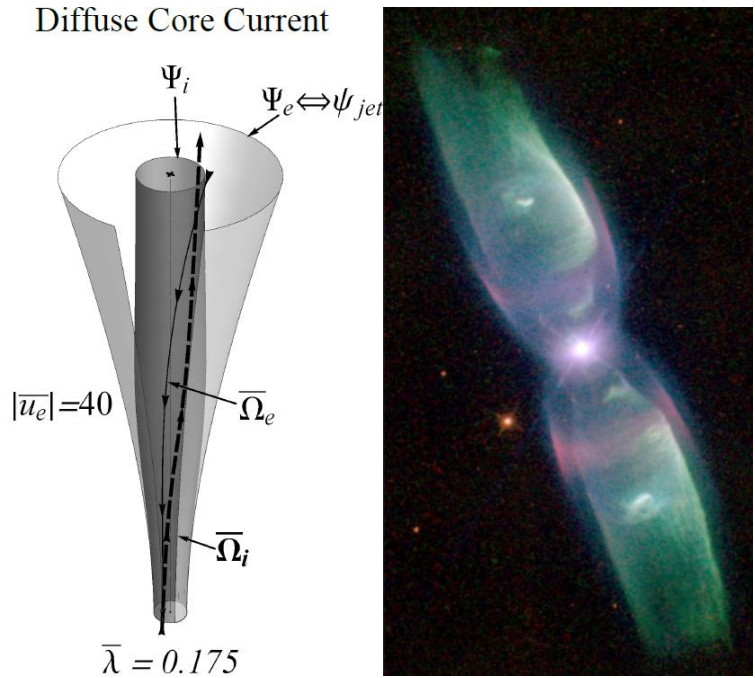

**Figure 7. (Left)** Bulged ion canonical vorticity flux tube $\Psi_i$ (dark gray) for a diffuse core current jet with low current ($\bar{\lambda} = 0.175$) and a large electron velocity ($|\bar{u}_e| = 40$). Equation (5) shows that density gradients can be normal to the dark gray canonical $\Psi_i$ surfaces rather than the light gray magnetic flux surfaces $\psi$ suggesting that ion line emission reveals canonical flux tubes as opposed to magnetic flux tubes. **(Right)** M2-9 Butterfly nebula (NASA HST STIS/CCD – MIRVIS), a bipolar planetary nebula with distinctive bulging that resembles the ion canonical flux tube in the left panel. Transitions in the geometry of the jet (i.e. flaring to collimation) may therefore also reflect changes in the relative strength of the fluid flow vorticity flux to the magnetic flux. Image credit: Bruce Balick (University of Washington), Vincent Icke (Leiden University), Garret Mellema (Stockholm University), and NASA

When $\bar{\omega}_{ir} < 0$ and $|\bar{\omega}_{ir}| > |\bar{B}_r|$, ion canonical flux surfaces can actually be pulled inward ($-\hat{r}$ direction) resulting in a bulged canonical flux tube that narrows before collimating at a constant final collimation radius $a_{\Psi_i}$ (Figure 7). This occurs for the diffuse core current jet at low $\bar{\lambda}$ and large $|\bar{u}_e|$ values, e.g. $\Psi_i$ contour in Figure 6(c) with $\bar{\lambda} = 0.5$ and $|\bar{u}_e| = 10$ where the bulge is visible at $\bar{z} \sim 20$ and $\bar{r} \sim 2.4$. As $\bar{\lambda}$ is increased and $\bar{\omega}_{iz}$ begins to dominate, any bulging or narrowing becomes less prominent and the final collimation radius $a_{\Psi_i}(\bar{l})$ actually becomes slightly larger again ($\Psi_i$ contours in Figure 6(c) with $\bar{\lambda} = 10$ with $a_{\Psi_i}(\bar{l}) \sim 1.8$ versus $\bar{\lambda} = 2$ with $a_{\Psi_i}(\bar{l}) \sim 1.4$). For both current profiles, the final collimation radius appears to approach a constant value with increasing $\bar{\lambda}$ and $|\bar{u}_e|$ values, i.e. for a fixed $|\bar{u}_e|$ the final collimation radius approaches a constant and will not change further with increasing $\bar{\lambda}$; and for a fixed $\bar{\lambda}$, the final collimation radius approaches a different constant with increasing $|\bar{u}_e|$. For the case of the uniform current profile, the approach toward smaller radii is monotonic with increasing $\bar{\lambda}$ and $|\bar{u}_e|$. For the diffuse core current, however, the approach is monotonic with $|\bar{u}_e|$ (Figure 6(d)), but not with $\bar{\lambda}$ (Figure 6(c)) where a minimum collimation radius $a_{\Psi_i}(\bar{l}) \sim 1.4$ is observed at $\lambda \sim 2$. Since (in reality) ions should be tied to canonical vorticity flux surfaces (equations (4) and (5)), any light emission from ions would be signatures of canonical vorticity flux tubes rather than magnetic flux surfaces, with any associated bulges and collimation dependent on the current and flow profiles (and because of the massless electron assumption here, electron emission would be signatures of magnetic flux surfaces).

### *4.2 Diffuse-core-with-skin and Skin-only Current Profiles*

Figure 8 presents the morphology of a flared plasma jet with diffuse-core-with-skin and skin-only current profiles from the MHD perspective of current-carrying magnetic flux tubes at low and high $\bar{\lambda}$ with a bulk velocity $\bar{\boldsymbol{U}}$ also effected by the choice of $|\bar{\boldsymbol{u}}_e|$. Again, as $\bar{\lambda}$ is increased, the magnetic field becomes helical in regions that enclose a net current. For the diffuse-core-with-skin current profile, this occurs throughout the jet with a shear due to gradients in current density profile $j_z(r)$. However, for the case of the skin current, the magnetic field in the core of the jet ($\eta < 0.6$) remains untwisted because of the absence of current in this region. This results in significant shear in the skin region because $B_\theta$ must increase over a shorter distance to match the final value at the boundary to satisfy Ampère's law. Bulk flow separation from magnetic field streamlines is again apparent in regions of finite current density, with lower electron velocities resulting in increasing pitch angle. Because there is no toroidal current ($j_\theta = 0$), magnetic flux surfaces remain unchanged as the current is increased. From a canonical flux tube point-of-view (Figure 9), the shape of canonical flux surfaces depends on the species flow velocity as well as on jet current and current profile (Figure 10). Figures 9(a)-(d) show the equivalent systems to Figures 8(a)-(b) (diffuse-core-with-skin profile) for the same electron flow velocities used in section 4.1, while Figures 9(e)-(h) show equivalent systems to Figures 8(c)-(d) (skin current profile). Again the electron canonical vorticity flux tubes are geometrically coincident to the magnetic flux tubes $\Psi_e \Leftrightarrow \psi_{jet}$ because of our massless electron assumption, while the ion canonical vorticity flux surfaces depart magnetic flux surfaces $\Psi_i \Leftrightarrow \psi_{\text{jet}}$ (in regions of finite $\boldsymbol{\omega}_{\text{ipol}}$) to become increasingly collimated further up away from the footpoint. For the diffuse-core-with-skin current profile, collimation is complete in regions closer to the footpoint at larger currents $\bar{\lambda}$ and electron flow velocities $|\bar{\boldsymbol{u}}_e|$, whereas for the skin current profile, collimation is never complete due to the absence of current in the core region of the jet (Figure 10).

Similar to section 4.1, for both current profiles $\bar{\omega}_{iz} \geq 0$ everywhere since the total current increases monotonically with radius. For the diffuse-core-with-skin current, the radial profile of $\bar{\omega}_{iz}(\bar{r})$ is peaked on axis with a secondary local maximum at $\eta = 0.8$ (i.e. peak of the skin current). For identical jet parameters $\bar{\lambda}$ and $|\bar{\boldsymbol{u}}_e|$, the magnitude of $\bar{\omega}_{iz}$ in the core region falls between that of the uniform and diffuse core current jets of section 4.1. However, due to the contribution of a peaked skin current, the magnitude of $\omega_{iz}$ in the skin region exceeds that of the uniform and diffuse core current profiles for similar jet parameters. As a result, the narrowing and bulging effect observed for the diffuse-core current jet at low $\bar{\lambda}$ and high $|\bar{\boldsymbol{u}}_e|$ (Figure 6(c)) is less pronounced even though a finite, negative $\bar{\omega}_{ir}$ is observed near the jet boundary (Figure 10(a)). A transition within the skin region of $\bar{\omega}_{ir}$ from negative to positive (moving in the inward radial direction) further restricts any narrowing of the canonical flux tube. Consequently, for identical jet parameters, collimation of the diffuse-core-with-skin current jet occurs closer to the footpoint than for uniform current, but farther away than for the purely diffuse core current jet. The approach toward tighter radii is monotonic with increasing $\bar{\lambda}$ and $|\bar{\boldsymbol{u}}_e|$ (Figures 10(a)-(b)) as was the case for the uniform current profile (Figures 6(a)-(b)).

For the skin current profile, the absence of current in the core region of the jet prevents collimation of the ion canonical flux tube. With no poloidal current in this region, magnetic fields are untwisted as are the species velocity fields; consequently, $\boldsymbol{\omega}_{\text{ipol}} = 0$ and the ion canonical flux surfaces $\Psi_{i,\text{inner}}$ are equivalent to magnetic flux surfaces $\psi_{\text{inner}}$. In the skin region, however, a large radial current gradient results in a strongly peaked $\bar{\omega}_{iz}(\bar{r})$ profile that works to collimate canonical flux surfaces $\Psi_i$. The presence of a negative $\bar{\omega}_{ir}$ at the jet boundary further helps to separate canonical flux surfaces, but any narrowing of the flux tube is again constrained by a transition (in the inward radial direction) of $\bar{\omega}_{ir}(r)$ to positive values (Figure 10(c)). As $\bar{\lambda}$ and $|\bar{\boldsymbol{u}}_e|$ are increased, canonical flux surfaces approach the current free region of the jet but can go no further (Figures 10(c)-(d)). Because the current is self-similar along the magnetic flux tube, the current free region is flared and therefor the ion canonical flux tubes can never fully collimate but must follow the less-pronounced flaring of the current free region. The shell between the ion canonical flux tube (e.g. ion line emission) and the magnetic flux tube (e.g. electron synchrotron emission) then represents the skin current density shell.

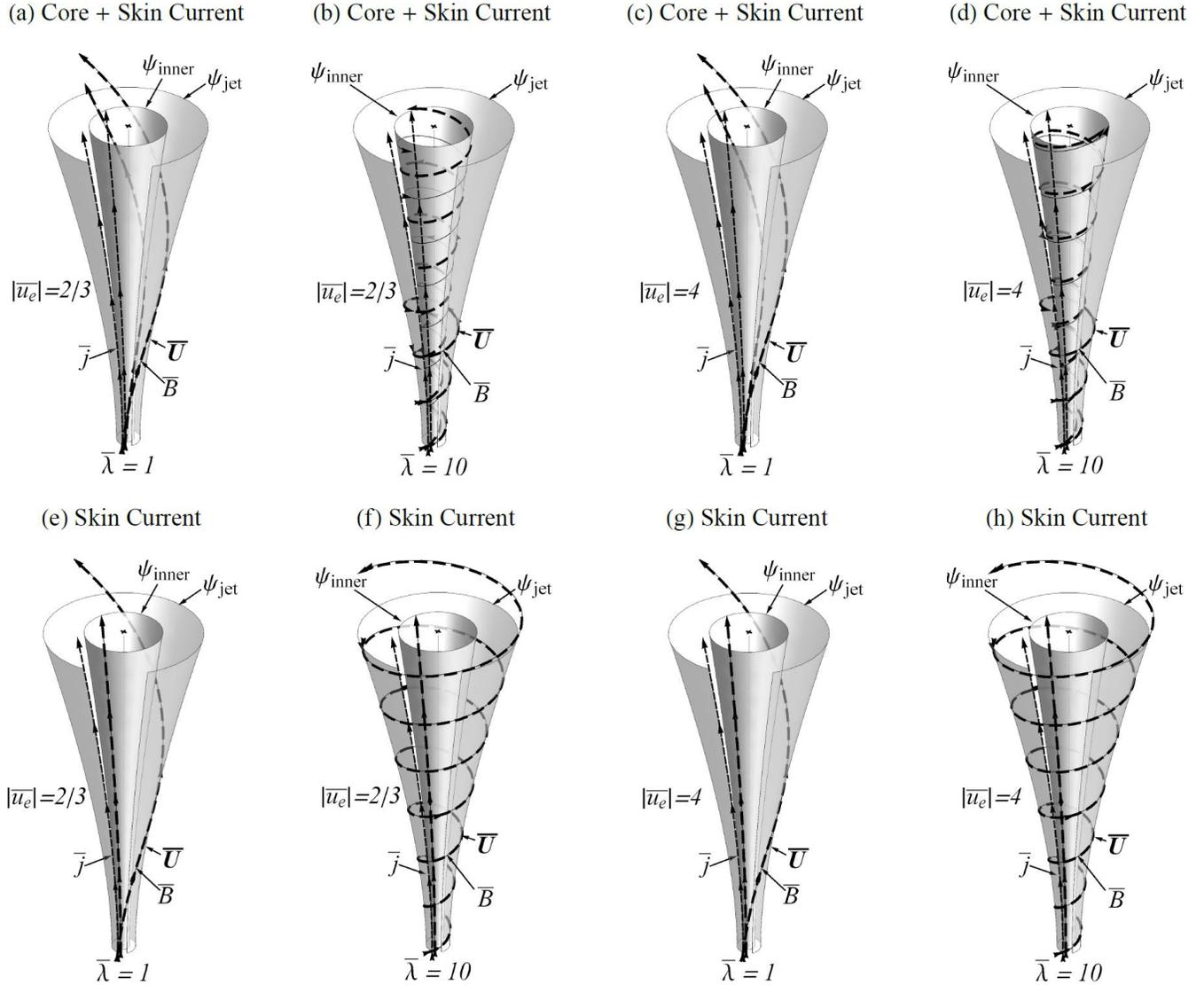

**Figure 8.** Flared magnetic flux tubes with diffuse-core-with-skin and skin-only current profiles for sub-Alfvénic ($|\bar{u}_e| = 2/3$) and super-Alfvénic ($|\bar{u}_e| = 4$) electron flow velocities at low $\bar{\lambda} = 1$ and high $\bar{\lambda} = 10$. Thin solid lines are magnetic field streamlines ($\bar{B}$), thin dashed lines are current density streamlines ($\bar{j}$), and bold dashed lines are bulk flow velocity streamlines ($\bar{U}$). Arrows on the streamline indicate the direction of the vector field. Streamlines on both the outer magnetic flux surface ($\psi_{jet}$) and an inner magnetic flux surface ($\psi_{inner}$) are shown; however, some streamlines on the outer magnetic flux surface have been truncated for visual clarity. Figure 5 shows the same systems from a canonical flux tube point-of-view.

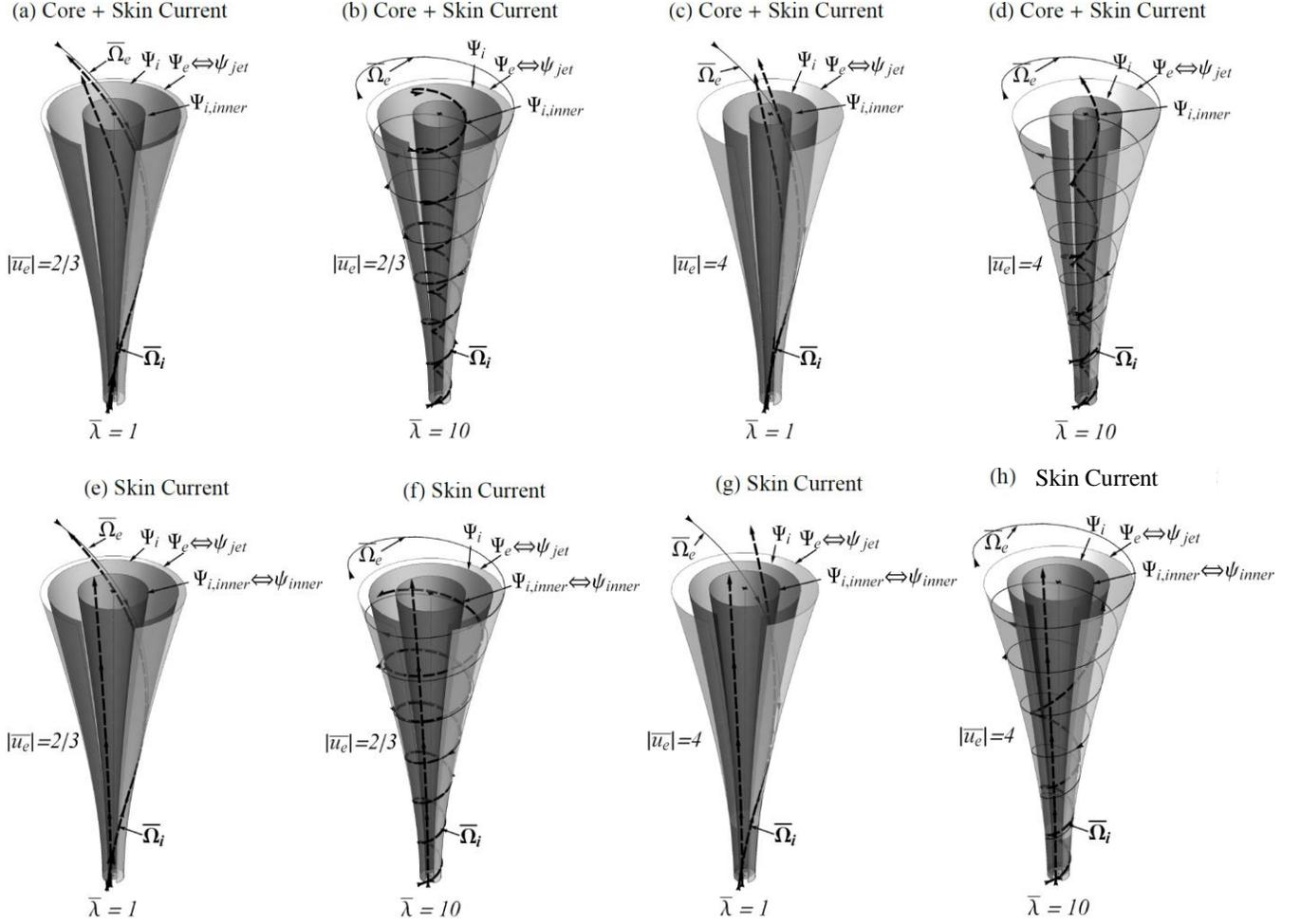

**Figure 9.** Canonical vorticity flux tubes with diffuse-core-with-skin and skin-only current profiles for sub-Alfvénic ($|\bar{u}_e| = 2/3$) and super-Alfvénic ($|\bar{u}_e| = 4$) electron flow velocities at low $\bar{\lambda} = 1$ and high $\bar{\lambda} = 10$. These canonical flux tubes are equivalent to the corresponding magnetic flux tubes of Figure 8. Ion canonical vorticity streamlines ($\bar{\boldsymbol{\Omega}}_i$) are drawn as bold, dashed lines along two ion canonical vorticity flux tubes ($\Psi_i$ and $\Psi_{i,inner}$), while electron canonical momentum ($\bar{\boldsymbol{\Omega}}_e$) streamlines (geometrically equivalent to magnetic field streamlines), are depicted as a thin, solid line along an electron canonical flux tube ($\Psi_e \Leftrightarrow \psi_{jet}$, light gray tube). Some streamlines on the outer ion canonical vorticity flux tube ($\Psi_i$, dark gray tubes) have been truncated for visual clarity. As the current is increased, ion canonical flux tubes depart from magnetic flux tube surfaces and begin to collimate. For the skin current (e-h), collimation never completes due to the absence of poloidal flow vorticity flux in the core of the jet but the flaring is less pronounced than the flare of the magnetic flux tube.

## 5. KINEMATIC EVOLUTION OF RELATIVE CANONICAL HELICITY

Figure 11 shows the kinematic evolution of the normalized (weighted) ion relative canonical helicity components (magnetic helicity $\bar{\mathcal{K}}$, cross-helicity $\bar{\mathcal{X}}_i$, kinetic helicity $\bar{\mathcal{H}}_i$ components of the canonical helicity $\bar{K}_{irel}$ in equation (24)) for all cases considered in Section 4 as $\bar{\lambda}$ is increased from $10^{-4}$ to 40. Integrations are performed using a trapezoidal method with a precision of $\pm 1\%$. The normalized helicity values have been transformed using a logarithmic modulus transformation $K' = \text{sign}(K) \log_{10}(|K| + 1)$ that preserves any zero crossings. The electron canonical helicity, although not shown explicitly, is equivalent to the weighted magnetic helicity component because of our massless electron assumption.

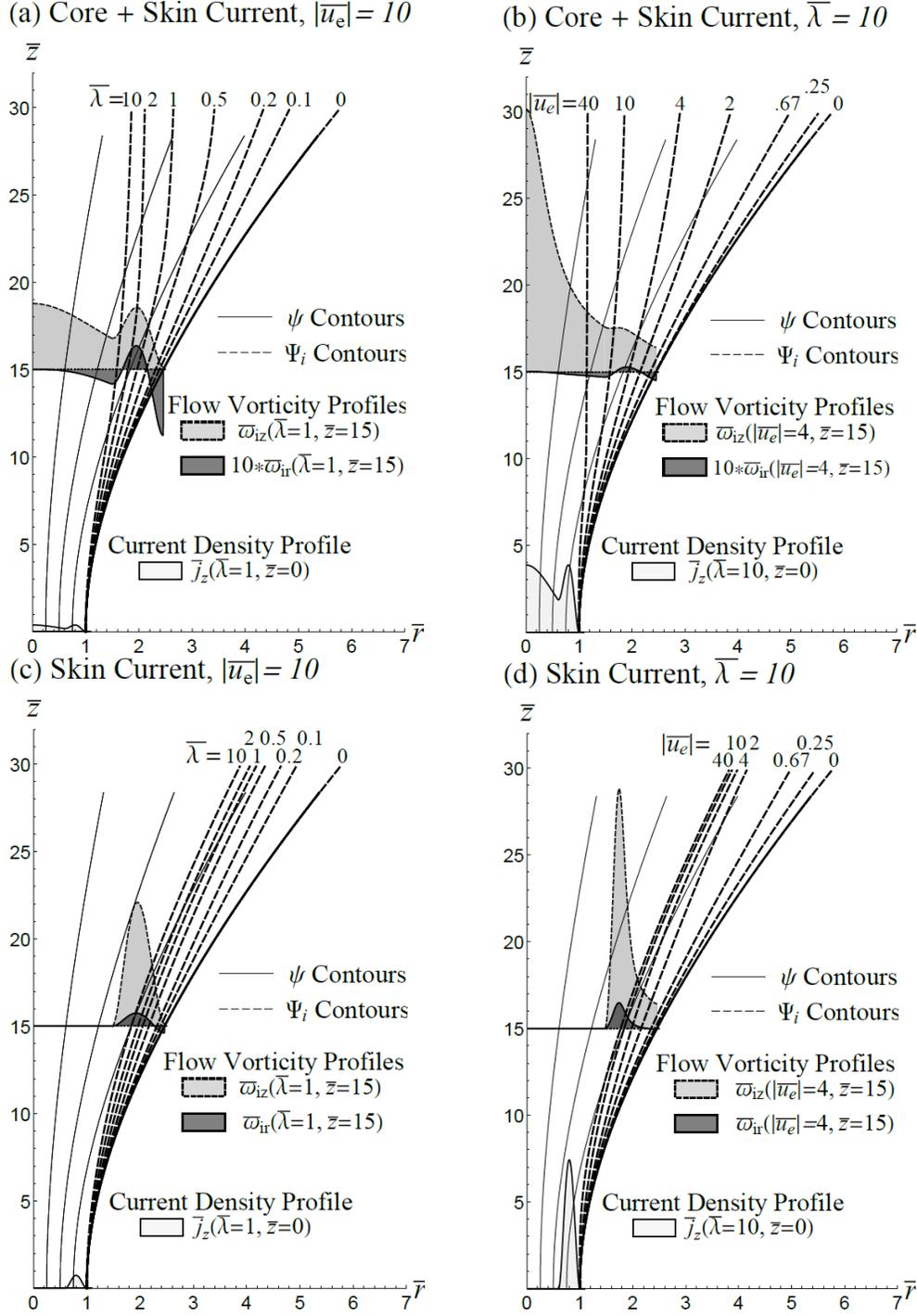

**Figure 10.** Dependence of ion canonical vorticity flux surface $\Psi_i$ on total jet current $\bar{\lambda}$ (panels a, c) and on electron fluid velocity $|\bar{u}_e|$ (panels b, d) for diffuse-core-with-skin (panels a, b) and skin-only (panels c, d) currents. For both current profiles, canonical flux surfaces separate closer to the base of the jet with increasing jet current $\bar{\lambda}$ and electron velocity $|\bar{u}_e|$; however, for the skin-only current, collimation does not occur due to the absence of poloidal flow vorticity flux in the core of the jet. Plots of the current profile $\bar{j}_z(\bar{r})$ are shown at the base of the jet (lightest gray) while representative poloidal flow vorticity profiles $\bar{\omega}_{ir}(\bar{r})$ (dark gray) and $\bar{\omega}_{iz}(\bar{r})$ (light gray) are drawn at $\bar{z}=15$. Contours of constant magnetic flux $\psi$ (solid lines) do not change with $\bar{\lambda}$ nor $|\bar{u}_e|$ and are presented to reveal passage of the canonical flux surface through different regions of the current profile.

These plots demonstrate the influence of fluid momentum and current profile on canonical helicity, while revealing distinct regions in $\bar{\lambda}$ space where different helicity components dominate.

For the current profiles and electron velocities considered in section 4, cross and magnetic helicity dominate the total helicity content of the jet (Figure 11). At low $\bar{\lambda}$, the cross-helicity dominates because ions flow roughly along the magnetic field lines with a toroidal vorticity parallel to the toroidal magnetic vector potential. As $\bar{\lambda}$ increases, magnetic helicity increases sufficiently rapidly to dominate cross-helicity above a threshold $\bar{\lambda}$. The threshold where this interchange occurs is around $0.001 < \bar{\lambda} < 0.01$. For cases with larger flow velocities ($|\bar{\boldsymbol{u}}_e| = 4$ in Figure 11(d), (f), (h)), as $\bar{\lambda}$ is increased above another threshold around $\bar{\lambda} \sim 2$, cross-helicity once again dominates magnetic helicity due primarily to a large poloidal flow vorticity and poloidal magnetic vector potential.

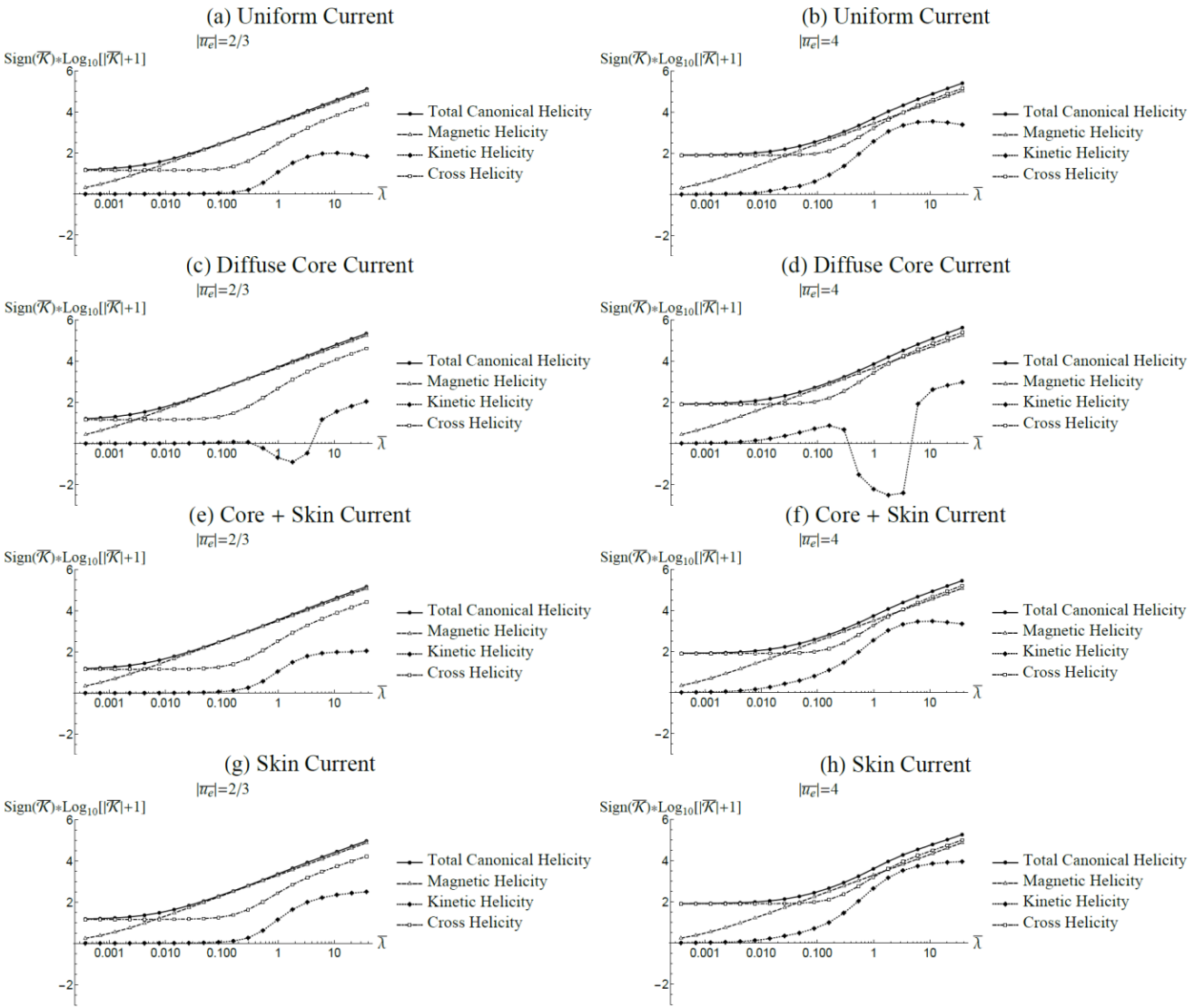

**Figure 11.** Kinematic evolution of the weighted normalized ion relative canonical helicity components for select current profiles and electron flow velocities. For all the cases considered in Section 4 cross and magnetic helicity components dominate the total canonical helicity content of the jet. Ion kinetic helicity increases at moderate values of $\bar{\lambda}$ but levels off or even decreases as the ion flows unwind at high $\bar{\lambda}$ (see the text, Figure 13). For the diffuse core current profile (panels c, d), a negative dip in the ion flow helicity is observed at moderate $\bar{\lambda}$ as a consequence of a changing $\bar{\omega}_{i\theta}(r)$ profile.

At larger electron velocities, the magnitude of cross and flow helicity increases while the overall shape of their curves is maintained. By increasing the electron velocity well above $|\bar{u}_e| = 4$, (i.e., approaching neutral fluid limit), scenarios emerge in which the ion flow helicity can dominate (followed by cross-helicity) over large regions of $\bar{\lambda}$ space. This is demonstrated in Figure 12, where the normalized electron velocity is set to $|\bar{u}_e| = 40$ for a uniform current jet. In this case, ion kinetic helicity grows with magnetic helicity at low $\bar{\lambda}$ and becomes the dominant helicity component between $0.2 \lesssim \bar{\lambda} \lesssim 10$. In regions outside of this range the cross-helicity dominates all other helicity components.

In all cases, a reduction in the growth of ion flow helicity $\bar{\mathcal{H}}_i$ is seen at high $\bar{\lambda}$. This can be explained by noting that, at all locations where the current density is large, ions must compensate for any toroidal electron flow and increase their pitch angle (i.e. untwist). This unravelling of ion flow tubes is illustrated in Figure 13 for the case of a uniform current jet with sub-Alfvénic electron velocities. Because ion flows only unwind in regions of finite current density, the impact on the ion flow helicity $\bar{\mathcal{H}}_i$ is most pronounced for jets in which the current density is more uniformly distributed throughout the current-carrying region. In fact, for the case of the uniform current and diffuse-core-with-skin current profiles, the ion flow helicity is observed to decrease for $\bar{\lambda} \gtrsim 10$ whereas for the diffuse-core current and skin current profiles, the decrease in growth rate is less severe.

Another interesting feature is the negative dip in ion flow helicity for the case of the diffuse core current (Figure 11 (c)-(d)). To explain this, the ion relative flow helicity can be expanded as

$$\bar{\mathcal{H}}_{irel} = \oiiint \bar{u}_{ir}(\bar{\omega}_{ir} + \bar{\omega}_{ir,ref}) + (\bar{u}_{i\theta} - \bar{u}_{i\theta,ref})\bar{\omega}_{i\theta} + \bar{u}_{iz}(\bar{\omega}_{iz} + \bar{\omega}_{iz,ref})d\bar{V}. \quad (28)$$

The reference fields are obtained from equation (26) and are generally opposite in sign to the actual fields as a result of the boundary conditions enforced to ensure gauge independence. For the diffuse core current, the first term in the integrand is small and negative over the $\bar{\lambda}$ range since $\bar{\omega}_{ir}$ is generally negative (Figure 6 (c)-(d)) because of the shape of the current profile and flaring of the jet ($u_{ir}$ is always positive or zero). The reference field $\bar{\omega}_{ir,ref}$ in this case does not fully cancel $\bar{\omega}_{ir}$ when integrated over the jet volume. Because the axial flow velocity $\bar{u}_{iz}$ and vorticity $\bar{\omega}_{iz}$ are always positive, the third term in the integrand is always positive (again the reference field does not fully cancel the actual field upon integration). This term grows with increasing $\bar{\lambda}$ as the magnitude of $\bar{u}_{iz}$ and $\bar{\omega}_{iz}$ increases within the jet. The second term in the integrand varies most due to the complicated dependence of $\bar{\omega}_{i\theta}$ on the current profile and flaring angle of the jet. Using equation (17) to expand the toroidal vorticity in terms of the electron velocity and current density

$$\bar{\omega}_{i\theta} = \frac{\partial}{\partial \bar{z}}(\bar{J}_r + \bar{u}_{er}) - \frac{\partial}{\partial \bar{r}}(\bar{J}_z + \bar{u}_{ez}) \quad (29)$$

it can be seen that at zero jet current ($\partial \bar{J}_r/\partial \bar{z} = \partial \bar{J}_z/\partial \bar{r} = 0$), $\bar{\omega}_{i\theta}$ is finite and positive because electrons flow with a constant velocity along flared magnetic field lines. The radial profile $\bar{\omega}_{i\theta}(\bar{r})$ peaks at the wall of the jet; however, as the current is increased, the toroidal ion flow vorticity decreases in the outer region as a consequence of the negative gradient in the toroidal magnetic field (i.e. increasing $|\bar{u}_{ez}|$ with $\bar{r}$) while increasing in the core due to the negative radial gradient in current density and positive gradient in the toroidal magnetic field (i.e. decreasing $|\bar{u}_{ez}|$ with $\bar{r}$). Meanwhile, the first term on the right of equation (29) remains negative and small compared to the second because of the shallow flaring of the jet. As the current is increased, the volume where $\omega_{i\theta} < 0$ (and $\bar{u}_{i\theta}$ is large) begins to exceed that where $\omega_{i\theta} > 0$ and term 2 in equation (28) becomes negative. This term grows faster in magnitude than term three and causes the ion flow helicity to dip negative around $\bar{\lambda} \sim 0.3$. Increasing the jet current beyond $\bar{\lambda} \sim 1.5$ and the toroidal ion flow vorticity becomes increasingly peaked in the core of the jet, causing the integral term to increase. This together with the growth of the third term in equation (28) is responsible for the negative-to-positive transition observed in the relative ion flow helicity at $\bar{\lambda} \sim 3$.

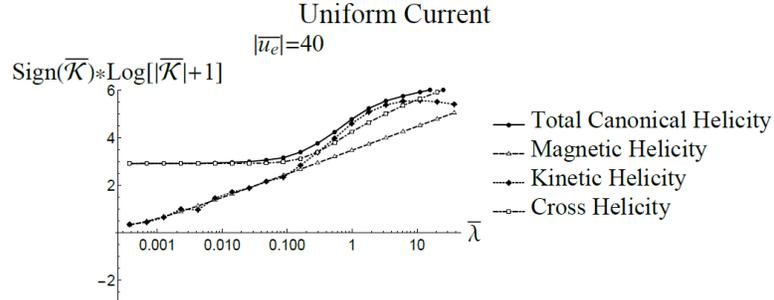

**Figure 12.** Kinematic evolution of the normalized ion relative canonical helicity components for a uniform current profile with a normalized electron flow velocity $|\bar{\boldsymbol{u}}_e| = 40$. In this case, the ion kinetic helicity dominates at moderate $0.2 \lesssim \bar{\lambda} \lesssim 10$, with cross-helicity dominating at larger $\bar{\lambda}$.

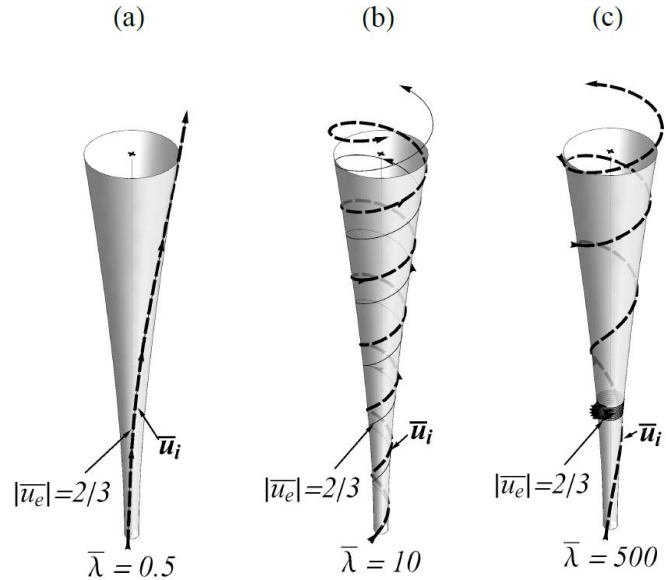

**Figure 13.** Ion flow tubes for the case of a uniform current profile with increasing jet current $\bar{\lambda}$. Ion flow streamlines (dashed line) twist as $\bar{\lambda}$ is increased (panel a to b) then unwind at even larger $\bar{\lambda}$ (panel b to c) to compensate for increasing electron twist (solid line) in carrying the poloidal current. This behavior explains the reduction in ion flow helicity observed at large $\bar{\lambda}$ values for current profiles that are more distributed throughout the cross-section of the jet (Figure 11).

## 6. SUMMARY

We have presented a more fundamental framework for examining magnetized flows in a jet geometry by examining the geometry and topology of canonical momentum fields in a flared current-carrying magnetic flux tube. In essence, this single framework performs a transformation from the familiar equations of motion to a generalized Maxwell equations field formulation (You 2016), to demonstrate that the canonical flux tube geometry point-of-view is valid in regimes beyond the fluid regime and, in particular, can be applied to kinetic and relativistic regimes. The self-consistent dynamics at kinetic regimes is the subject of ongoing work, while the stationary reduced two-fluid cases are considered here.

By varying the enthalpy boundary conditions (i.e. current profile and flow velocity), we reveal the dependence of canonical fields on the underlying magnetic and flow components and show that canonical flux tubes can become long and collimated even when the magnetic field is flared. For the current profiles considered, ion canonical flux tubes collimate as a result of positive flow vorticity flux. The final radius and location at which the collimation occurs was shown to depend on the total jet current $\bar{\lambda}$, current profile $\bar{j}_z(r)$, and flow momentum $|\bar{u}_e|$. Taken with the fact that equation (4) generalizes the frozen-in magnetic flux condition to the more fundamental frozen-in canonical vorticity flux, our results suggest that collimated astrophysical jets in nature may trace canonical vorticity flux tubes as opposed to magnetic flux tubes. Transitions in the shape of such jets (i.e. flaring, bulging, and collimation) could therefore reflect variation in the relative strength of magnetic and fluid vorticity fluxes. The tendency of canonical flux tubes to collimate agrees with observations that some astrophysical jets become increasingly collimated further away from their origin (Ray & Mundt 1993).

By accounting for finite particle momentum, the canonical viewpoint preserves non-ideal effects that are lost in ideal MHD, while retaining intuition of the changing geometry of flux tubes. Concepts such as magnetic helicity conservation, which govern the evolution of magnetic flux tubes into other magnetic flux tubes, can therefore be generalized to canonical helicity conservation, which predicts the topological evolution of both magnetic and flow tubes. We calculated the gauge-invariant relative canonical helicity content of the jet and reveal the sensitivity of various (magnetic, cross, and kinetic) helicity components to the shape and magnitude of the current profile and electron flow velocity. At large jet currents, we observe a decrease in the ion kinetic helicity as ion flow fields unwind to compensate for increasing magnetic twist. We hypothesize that conversion between helicities at the appropriate scales [You 2012, 2014] could convert helical magnetic twist into helical shear flows, which could be sufficiently strong to stabilize against current-driven instabilities. Testing this hypothesis would require solving for the full dynamics of the helicity transport equation and canonical equation of motion, and verification in a laboratory astrophysical jet experiment, which are the subject of ongoing work and construction.

Acknowledgements

This work was supported by the U.S. Department of Energy Grant No. DE-SC0010340. The authors would like to thank B. Balick for the use of the image of M2-9 in Figure 7. Calculations and figures produced using Wolfram Research, Inc., Mathematica, Version 10.2, Champaign, IL (2015).

References

Avinash, K. 1992, PhFlB, 4, 3856.
Bellan, P. 2000, Spheromaks, (London: Imperial College Press).
Bellan, P. 2003, PhPl, 10, 5.
Bellan, P., You S., & Hsu S. Astrophys. Space Sci., 298, 203-209, 2005
Broderick, A., & Loeb A. 2009, ApJ, 703, 104.
Brown, M. 1997, PlPh, 57, 1.
C. Burrows, K. Stapelfeldt, A. Watson, et al. 1996, ApJ, 473, 1.
DeYoung, D. 1991, Sci, 252, 5004.
Donati,J., Howarth, I., Jardine, M.,Petit, P., et al. 2006, MNRAS, 370, 629.
Finn, J. M., & Antonsen, T. M. 1985, CoPPC, 9, 111.
Fiksel, G., Almagri, A. F., Chapman, B. E., et al. 2009, PhRvL, 103, 145002.

Frank, A., Ryu, D., Jones, T. W., & Noriega-Crespo, A. 1998, ApJ, 494, 79.
Icke, V. Mellema, G. Balick, B. Eulerink, F., & Frank, A.1992, Natur, 355, 6360.
Jarboe, T. R., Hamp, W. T., Marklin, G. J., et al. 2006, PhRvL, 97, 115003.
Ji, H. 1999, in Magnetic Helicity in Space and Laboratory Plasmas, ed M. Brown et al. (Washington, DC: American Geophysical Union), 167.
Kawamori, E. Murata, Y. Hirota, D., et al. 2005, NucFu, 45, 843.
D. Lynden-Bell, 2003. MNRAS., 341, 4.
Moffat, H. 1969, JFM, 35, 117.
M. Nakamura, Y. Uchida, & S. Hirose. 2001, NewA, 6, 2.
Norman, M., Smarr, L., Winkler, K. & Smith, M. 1982, A&A, 113, 2.
Oliveira, S. R., & Tajima, T. 1995, PhRvE, 52, 4287.
Ray, T. P., & Mundt, R. 1993, in Astrophysical Jets, ed D. Burgarella et al. (Caimbridge: University Press), 149.
R. Sahai, M. Morris, G. Knapp, K. Young, & C. Barnbaum. 2003, Natur, 426, 6964.
Shumlak, U., & Hartman, C. 1995, PhRvL, 75, 18.
Shumlak, U. Nelson, B., Golingo, R., et al. 2003, PhPl, 10, 5.
Steinhauer, L., & Ishida, A. 1997, PhRvL, 79, 3423.
Steinhauer, L., & Ishida, A. 1998, PhPl, 5, 2609.
Steinhauer, L., Yamada, H., & Ishida, A. 2001, PhPl, 8, 4053.
Taylor, J. B. 1974, PhRvL, 33, 1139.
Taylor, J. B. 1986, RvMP, 58, 741
Turner, L. 1985, ITPS 14, 849.
Woltjer, L. 1986, PNAS, 44, 833
You. S. 2012, PhPl, 19, 092107.
You, S. 2014, PPCF, 56, 064007.
You, S. 2016, PhPl, 23, 072108.
You, S., Yun, G., & Bellan, P. 2005, PhRvL, 95, 4.
Yun, G., You, S., & Bellan, P. 2007, NucFu, 47, 3.